\documentclass[11pt,a4paper]{article}
\usepackage{jcappub}

\usepackage{pdflscape}
\usepackage{amsmath}
\usepackage{amssymb}
\usepackage{dcolumn}
\usepackage{bm}
\usepackage{color}
\usepackage{epsfig}
\usepackage{amsfonts}
\usepackage{graphicx}
\usepackage{subfigure}
\usepackage{dcolumn}

\newcommand{\be}{\begin{equation}}
\newcommand{\ee}{\end{equation}}
\newcommand{\bea}{\begin{eqnarray}}
\newcommand{\eea}{\end{eqnarray}}

\def\bx{{\bf x}}
\def\bk{{\bf k}}

\begin{document}

\title{$f(T,\mathcal{T})$ gravity and cosmology}

\author[a]{Tiberiu Harko}
\author[b]{Francisco S. N. Lobo}
\author[c]{G. Otalora}
\author[d,e]{Emmanuel N. Saridakis}

\affiliation[a]{Department of Mathematics, University College London, Gower
Street, London
WC1E 6BT, United Kingdom}

\affiliation[b]{Centro de Astronomia e Astrof\'{\i}sica da Universidade de
Lisboa, Campo
Grande, Edific\'{\i}o C8, 1749-016 Lisboa, Portugal}

\affiliation[c]{Departamento de F\'{\i}sica, ICE, Universidade Federal de Juiz
de Fora, Caixa Postal 36036-330, Minas Gerais, Brazil}

\affiliation[d]{Physics Division, National Technical University of Athens,
15780 Zografou Campus,  Athens, Greece}

\affiliation[e]{Instituto de F\'{\i}sica, Pontificia Universidad  Cat\'olica
de Valpara\'{\i}so, Casilla 4950, Valpara\'{\i}so, Chile}

\emailAdd{t.harko@ucl.ac.uk}
\emailAdd{flobo@cii.fc.ul.pt}
\emailAdd{gotalora@fisica.ufjf.br}
\emailAdd{Emmanuel$_-$Saridakis@baylor.edu}

\abstract{
We present an extension of $f(T)$ gravity, allowing for a general coupling
of the torsion scalar $T$ with the trace of the matter energy-momentum
tensor $\mathcal{T}$. The resulting $f(T,\mathcal{T})$ theory is a
new modified gravity, since it is different from all the existing torsion or
curvature based constructions. Applied to a cosmological framework, it leads
to interesting phenomenology. In particular, one can obtain a unified
description of the initial inflationary phase, the subsequent
non-accelerating, matter-dominated expansion, and then the transition to a
late-time accelerating phase. Additionally, the effective dark energy sector
can be quintessence or phantom-like, or exhibit the phantom-divide
crossing during the evolution. Moreover, in the far future the universe
results either to a de Sitter exponential expansion, or to eternal power-law
accelerated expansions.  Finally, a detailed study of the scalar
perturbations at the linear level reveals that $f(T,\mathcal{T})$ cosmology
can be free of ghosts and instabilities for a wide class of ansatzes and
model parameters. }


\maketitle

\section{Introduction}
\label{Introduction}

The verification of the late-time acceleration of the universe (see \cite{acc} for a detailed discussion of the recent astronomical observations) has led to
extensive research towards its explanation. This result  is based on
fitting a Friedman-Robertson-Walker type geometry, together with the corresponding cosmology,  to the existing astronomical data.
However, strictly speaking,
taking into account the present day astronomical observational information, the
only model-independent conclusion that we can infer at this stage  is that the
observations do not favor the pressureless Einstein-de Sitter model.

In general, there are two main
ways to achieve the goal of the theoretical explanation of the accelerated expansion of the universe.
The first direction consists in modifying the universe content, by
introducing a dark energy sector, starting either with a canonical scalar field, a
phantom field, or the combination of both fields in a unified model, and
proceeding to more complicated constructions (for reviews
see \cite{Copeland:2006wr,Cai:2009zp} and references  therein). 
The second
direction is to modify the gravitational sector itself (see
\cite{Capozziello:2011et,DeFelice:2010aj,Nojiri:2010wj,Lobo:2008sg} for reviews and references therein).
However, we mention that, up to physical interpretation issues, one can
transform, completely or partially, from one approach to the other, since the
important issue is the number of extra degrees of freedom  (for such a
unified point of view see \cite{Sahni:2006pa}). Thus, one could also have
combinations of both directions, in scenarios with various couplings between
gravitational and non-gravitational sectors.

In modified gravitational theories one usually generalizes the
Einstein-Hilbert action of General Relativity, that is, one starts from the
curvature description of gravity. However, a different and interesting class
of modified gravity arises when one extends the action of the equivalent
formulation of GR based on torsion. As it is known, Einstein constructed
also the ``Teleparallel Equivalent of General Relativity'' (TEGR)
in which the gravitational field is described by the torsion tensor and not
by the curvature one
\cite{Unzicker:2005in,TEGR,TEGR22,Hayashi:1979qx,JGPereira,Arcos:2005ec,Maluf:2013gaa}
(technically this is achieved by using the Weitzenb{\"{o}}ck connection
instead of the torsion-less Levi-Civita one). Then, the corresponding Lagrangian
given by the torsion scalar $T$, results from contractions of the torsion
tensor, like the Einstein-Hilbert Lagrangian $R$ results from contractions of
the curvature (Riemann) tensor. Thus, instead of starting from GR, one can
start from TEGR and construct the $f(T)$ modified gravity, by extending $T$
to an arbitrary function in the Lagrangian
\cite{Ferraro:2006jd,Ben09,Linder:2010py}. The interesting feature is that
although TEGR is completely equivalent with General Relativity at the level
of equations, $f(T)$ is different than $f(R)$ gravity, that is they form
different gravitational modifications. Hence, $f(T)$ gravity has novel
and interesting cosmological implications
\cite{Linder:2010py,Chen:2010vaoooo,Dent:2011zz,Chen:2010va,Zheng:2010am,Bamba:2010wb,Cai:2011tc,Sharif001,Li:2011rn,
Capozziello:2011hj,Daouda:2011rt,Wu:2011kh,Wei:2011aa,Atazadeh:2011aa,Farajollahi:2011af,Karami:2012fu,
   Iorio:2012cm,Cardone:2012xq,Jamil:2012ti,Bohmer:2011si,Ong:2013qja,Amoros:2013nxa,Nesseris:2013jea,Bamba:2013ooa,Basilakos:2013rua,Otalora:2014aoa,Paliathanasis:2014iva,Nashed:2014uta,
   Bengochea001,Wang:2011xf,Miao003,Boehmer:2011gw,Daouda001,Ferraro:2011ks,Gonzalez:2011dr,Capozziello:2012zj,Atazadeh:2012am}.
   Additionally,
note that if one
starts from TEGR, but instead of the $f(R)$ is inspired by
higher-curvature modifications of General Relativity, one can construct
higher-order torsion gravity, such as the $f(T,T_G)$ paradigm
\cite{Kofinas:2014owa,Kofinas:2014aka}, which also presents interesting cosmological
behavior. Finally, another modification of TEGR is to extend it inserting
the Weitzenb\"{o}ck condition in a Weyl-Cartan geometry via a Lagrange
multiplier, with interesting cosmological implications
\cite{WC1WC2,WC2}.

Nevertheless, in usual General Relativity one could proceed to modifications
in which the geometric part of the action is coupled to the non-geometric
sector. The simplest models are those with non-minimally coupled
\cite{Uzan:1999ch,deRitis:1999zn,Bertolami:1999dp,Faraoni:2000wk} and non-minimal-derivatively coupled
\cite{Amendola:1993uh,Capozziello:1999xt,Daniel:2007kk,Saridakis:2010mf,Sadjadi:2010bz}
scalar fields, but one could further use arbitrary
functions of the kinetic and potential parts such as in K-essence
\cite{ArmendarizPicon:2000ah},
resulting in the general Horndeski \cite{Horndeski} and generalized Galileon
theories \cite{DeFelice:2010nf,Deffayet:2011gz,DeFelice:2011bh}. However, since there is no theoretical
reason against couplings between the gravitational sector and the standard
matter one, one can consider modified theories where the matter Lagrangian is
coupled to functions of the Ricci scalar
\cite{Bertolami:2007gv,Bertolami:2008zh,Bertolami:2008ab,Bertolami:2009ic}, and extend the theory to arbitrary
functions $\left(R,\mathcal{L}_m\right)$  \cite{Harko:2008qz,Harko:2010mv,Harko:2012hm,Wang:2012rw,Harko:2014gwa}. Alternatively,
one can consider models where the Ricci scalar is coupled with the trace of
the energy momentum tensor $\mathcal{T}$ and extend to arbitrary functions,
such as in $f(R,\mathcal{T})$ theory 
\cite{Harko:2011kv,Momeni:2011am,Sharif:2012zzd,Alvarenga:2013syu,Shabani:2013djy},
or even consider
terms of the form  $R_{\mu\nu}T^{\mu\nu}$ \cite{fRT,Odintsov:2013iba}. We stress that
the above modifications, in which one handles the gravitational and matter
sectors on equal footing, do not present any problem at the theoretical
level, and one would only obtain observational constraints due to
non-geodesic motion.

Having these in mind, one could try to construct the above extended coupled
scalar-field and coupled-matter modified gravities, starting not from
GR but from TEGR. The incorporation of non-minimally coupled scalar-torsion
theories was performed in 
 \cite{Geng:2011aj,Wei:2011yr,Geng:2011ka,Xu:2012jf,Otalora:2013dsa,Geng:2013uga,Otalora:2013tba,Sadjadi:2013nb,Kucukakca:2013mya}, 
 where a scalar field couples
non-minimally to the torsion scalar $T$. Similarly, in \cite{Harko:2014sja}
non-minimally matter-torsion theories were constructed, where the matter
Lagrangian is coupled to a second $f(T)$ function. We mention that both
these scenarios are different than the corresponding curvature ones, despite
the fact that uncoupled GR coincides with TEGR. They correspond to novel
modified theories, with a novel cosmological behavior.

In the present work, we are interested in constructing $f(T,\mathcal{T})$
gravity, that is, allowing for arbitrary functions of both the torsion scalar
$T$ and the trace of the energy-momentum tensor $\mathcal{T}$.
We emphasize that the resulting theory differs from $f(R,\mathcal{T})$ gravity,
in that it is a novel modified gravitational theory, with no curvature-equivalent, and its
cosmological implications prove to be very interesting.
Similar work has also been explored in \cite{Kiani:2013pba}, where the stability of the specific de Sitter solution, when subjected to homogeneous perturbations, was analyzed. Furthermore, the constraints imposed by the energy conditions were considered, and the parameter ranges of the proposed model were found to be consistent with the above stability conditions.
In this work, we consider more general cases. In particular, we
find late-time accelerated solutions, as well as initial inflationary phases,
followed by non-accelerating matter-dominated expansions, resulting to
a late-time accelerating evolution.

The plan of the manuscript is outlined as follows: In Section \ref{fTmodelbasic}, we
review the $f(T)$ gravitational modification. In Section \ref{ourmodel}, we
construct $f(T,\mathcal{T})$ gravity, and we apply it in a cosmological
framework. In Section \ref{Cosmologicalbehavior}, we analyze the cosmological
implications of two specific examples. Finally, Section \ref{Conclusions} is
devoted to the conclusions.

\section{$f(T)$ gravity and cosmology}
\label{fTmodelbasic}

We start with a brief review of $f(T)$ gravity. Throughout the manuscript,
we use Greek indices to span the coordinate space-time and Latin indices to
span the tangent space-time. The fundamental field is the vierbein
${\mathbf{e}_A(x^\mu)}$, which at each point $x^\mu$ of the space-time forms
an orthonormal basis for the tangent space, namely $\mathbf{e}
_A\cdot\mathbf{e}_B=\eta_{AB}$, where $\eta_{AB}={\rm diag} (1,-1,-1,-1)$.
Furthermore, in the coordinate basis we can express it in terms of components
as  $\mathbf{e}_A=e^\mu_A\partial_\mu$. Thus, the metric tensor can be
expressed as
\begin{equation}
\label{metrdef}
g_{\mu\nu}(x)=\eta_{AB}\, e^A_\mu (x)\, e^B_\nu (x).
\end{equation}
In the teleparallel gravitational formulation   (the vierbein components at
different points are ``parallelized'' and this is what is what is
represented by the appellation ``teleparallel'') one uses the Weitzenb\"{o}ck
connection
$\overset{\mathbf{w}}{\Gamma}^\lambda_{\nu\mu}\equiv e^\lambda_A\:
\partial_\mu
e^A_\nu$ \cite{Weitzenb23} which leads to zero curvature, and not the
Levi-Civita one which leads to zero torsion. Hence, the gravitational field
is described by the
 torsion tensor
\begin{equation}
\label{torsion2}
{T}^\lambda_{\:\mu\nu}=\overset{\mathbf{w}}{\Gamma}^\lambda_{
\nu\mu}-%
\overset{\mathbf{w}}{\Gamma}^\lambda_{\mu\nu}
=e^\lambda_A\:(\partial_\mu
e^A_\nu-\partial_\nu e^A_\mu).
\end{equation}
Additionally, we introduce the contorsion tensor
$K^{\mu\nu}{}_{\rho}\equiv-\frac{1}{2}\Big(T^{\mu\nu}{}_{\rho}
-T^{\nu\mu}{}_{\rho}-T_{\rho}{}^{\mu\nu}\Big)$, and the tensor
$
S_{\rho}{}^{\mu\nu}\equiv\frac{1}{2}\Big(K^{\mu\nu}{}_{\rho}
+\delta^\mu_\rho
\:T^{\alpha\nu}{}_{\alpha}-\delta^\nu_\rho\:
T^{\alpha\mu}{}_{\alpha}\Big)$. From the torsion tensor, one constructs
the torsion scalar and the respective teleparallel Lagrangian
\cite{TEGR,TEGR22,Hayashi:1979qx,JGPereira,Arcos:2005ec,Maluf:2013gaa}
\begin{equation}
\label{torsionscalar}
T\equiv\frac{1}{4}
T^{\rho \mu \nu}
T_{\rho \mu \nu}
+\frac{1}{2}T^{\rho \mu \nu }T_{\nu \mu\rho}
-T_{\rho \mu}{}^{\rho }T^{\nu\mu}{}_{\nu}.
\end{equation}
Thus, if $T$ is used in an action and one performs variation in terms
of the vierbeins, one extracts the same equations as with General Relativity.
That is why Einstein dubbed this theory ``Teleparallel Equivalent
of General Relativity'' (TEGR).

One can start from TEGR in order to construct various gravitational
modifications. In particular, one can extend $T$ to $T+f(T)$, resulting to
the so-called $f(T)$ gravity, where the action is given 
\begin{eqnarray}
\label{actionfT}
S = \frac{1}{16\pi G}\int d^4x e \left[T+f(T)\right],
\end{eqnarray}
with $e = \text{det}(e_{\mu}^A) = \sqrt{-g}$, $G$ the Newton's constant, and
setting the speed of light to one. It is clear that TEGR and thus General
Relativity is obtained when $f(T)=0$.  However, note that $f(T)$ differs from
$f(R)$ gravity, despite the fact that TEGR coincides with General
Relativity at the level of the equations.

The cosmological applications of $f(T)$ gravity can be investigated
incorporating the matter sector in the action. Thus, the latter is finally given by
\begin{eqnarray}
\label{actionmatter}
S= \frac{1}{16\pi G }\int d^4x e
\left[T+f(T)+\mathcal{L}_m\right],
\end{eqnarray}
where the matter Lagrangian is considered to correspond to a perfect fluid
with energy density and pressure $\rho_m$ and $p_m$, respectively (one could
include the  radiation sector too). Variation of the action
(\ref{actionmatter}) with respect to the vierbein leads to  the field
equations
\begin{equation}
\label{eomgeneral}
\left(1+f'\right)\,\left[e^{-1}\partial_{\mu}(ee_A^{\rho}S_{\rho}{}^{\nu\mu
}) -e_{A}^{\lambda}T^{\rho}{}_{\mu\lambda}S_{\rho}{}^{\mu\nu}\right]
 +
e_A^{\rho}S_{\rho}{}^{\nu\mu}\partial_{\mu}{T} f''+\frac{1}{
4} e_ {A}^{\nu}[T+f] = 4\pi Ge_{A}^{\rho}\overset
{\mathbf{em}}T_{\rho}{}^{\nu},
\end{equation}
where we denote $f'=\partial f/\partial T$ and $f''=\partial^{2} f/\partial
T^{2}$, while $\overset{\mathbf{em}}{T}_{\rho}{}^{\nu}$ stands for the
usual energy-momentum tensor.

Additionally, in order to obtain a flat Friedmann-Robertson-Walker (FRW)
universe
\begin{equation}
ds^2= dt^2-a^2(t)\,\delta_{ij} dx^i dx^j \,,
\end{equation}
where $a(t)$ is the scale factor, we consider
\begin{equation}
\label{veirbFRW}
e_{\mu}^A={\rm
diag}(1,a(t),a(t),a(t)).
\end{equation}
Thus, with this vierbein ansatz, the equations of motion (\ref{eomgeneral})
give rise to the modified Friedmann equations
\begin{eqnarray}\label{background1}
&&H^2= \frac{8\pi G}{3}\rho_m
-\frac{f}{6}-2H^2f'\\
\label{background2}
&&\dot{H}=-\frac{4\pi G(\rho_m+p_m)}{1+f'-12H^2f''},
\end{eqnarray}
respectively, where $H\equiv\dot{a}/a$ is the Hubble parameter, and the overdot denote the
$t$-derivatives. We mention that we have incorporated the useful relation
\begin{eqnarray}
\label{TH2}
T=-6H^2,
\end{eqnarray}
which holds for an FRW geometry, and which is determined from Eq. (\ref{torsionscalar}) using
Eq. (\ref{veirbFRW}).

\section{ $f(T,\mathcal{T})$ gravity and cosmology}
\label{ourmodel}

In this section, we present a novel theory of gravitational modification,
extending the previously described $f(T)$ gravity. In particular, apart
from an arbitrary function of the torsion scalar, we will also allow for an
arbitrary function of the trace of the energy momentum tensor.
Thus, we consider the action
\begin{equation}
S= \frac{1}{16\,\pi\,G}\,\int d^{4}x\,e\,\left[T+f(T,\mathcal{T})\right]%
+\int d^{4}x\,e\,\mathcal{L}_{m},
\label{action1}
\end{equation}%
where $f(T,\mathcal{T})$ is an arbitrary function of the torsion scalar $T$
and of the trace $\mathcal{T}$ of the matter energy-momentum tensor  $%
\overset{\mathbf{em}}{T}_{\rho}{}^{\nu}$, and  $\mathcal{L}_{m}$ is the
matter Lagrangian density. Hereinafter, and following the standard
approach, we assume that $\mathcal{L}_{m}$ depends only on the
vierbein and not on its derivatives.

Varying the action, given by Eq.~(\ref{action1}), with respect to the vierbeins yields the
field equations
\begin{eqnarray}
&&\left(1+f_{T}\right) \left[e^{-1} \partial_{\mu}{(e
e^{\alpha}_{A}
S_{\alpha}^{~\rho \mu})}-e^{\alpha}_{A} T^{\mu}_{~\nu \alpha} S_{\mu}^{~\nu
\rho}\right] 
+\left(f_{TT} \partial_{\mu}{T}+f_{T\mathcal{T}} \partial_{\mu}{%
\mathcal{T}}\right) e^{\alpha}_{A} S_{\alpha}^{~\rho \mu}
\notag \\
&&\ \ \ \ \ \ \ \ \ \ \ \ \ \ \ \ \ \ \ \ \ \ \ \ \ \ \ \ + e_{A}^{\rho}
\left(\frac{f+T}{4}\right)  -f_{\mathcal{T}} \left(\frac{e^{\alpha}_{A} \overset{\mathbf{em}}{T}%
{}_{\alpha}^{~~\rho}+p e^{\rho}_{A}}{2}\right)=4\pi G e^{\alpha}_{A}
\overset%
{\mathbf{em}}{T}_{\alpha}{}^{\rho},
\label{geneoms}
\end{eqnarray}
where $f_{\mathcal{T}}=\partial{f}/\partial{\mathcal{T}}$ and
$f_{T\mathcal{T%
}}=\partial^2{f}/\partial{T} \partial{\mathcal{T}}$.

In order to apply the above theory in a cosmological framework, we insert
as usual the flat FRW vierbein ansatz (\ref{veirbFRW}) into the
field equations \eqref{geneoms}, obtaining the modified Friedmann equations:
\begin{equation}
H^{2}=\frac{8\pi G}{3} \rho_m-\frac{1}{6}\left(f+12 H^{2} f_{T}\right)+f_{%
\mathcal{T}} \left(\frac{\rho_{m}+p_{m}}{3}\right),
\label{Friedmann1}
\end{equation}
\begin{equation}
\label{F2}
 \dot{H}=-4\pi G \left(\rho_m+p_m\right)-\dot{H} \left(f_{T}-12 H^{2}
f_{TT}\right)    -H \left(\dot{\rho}_{m}-3\,\dot{p}_{m}\right)
f_{T\mathcal{T%
}}-f_{\mathcal{T}} \left(\frac{\rho_{m}+p_{m}}{2} \right). 
\end{equation}
We mention that in the above expressions we have used that
$\mathcal{T}=\rho_{m}-3\,p_{m}$, which holds in the case of a perfect matter
fluid.

Proceeding, we assume that the matter component of the Universe
satisfies a barotropic equation of state of the form $p_{m}=p_{m}\left(\rho
_{m}\right)$, with $w_m=:p_m/\rho_m$ its equation-of-state parameter, and
$c_{s}^{2}=dp_{m}/d\rho _{m}$ the sound speed. Note that due to homogeneity
and isotropy, both $\rho_m$ and $p_m$ are function of $t$ only, and thus of
the Hubble parameter $H$. Thus, Eq.~(\ref{F2}) can be re-written
as
\begin{equation}
\dot{H}=-\frac{4\pi G\left( 1+f_{\mathcal{T}}/8\pi G\right) \left( \rho
_{m}+p_{m}\right) }{1+f_{T}-12H^{2}f_{TT}+H\left( d\rho _{m}/dH\right)
\left( 1-3c_{s}^{2}\right)f_{T\mathcal{T}} }.
\end{equation}

By defining the energy density and pressure of the effective dark energy
sector as
\begin{equation}
\rho _{DE}=:-\frac{1}{16\pi G}\left[ f+12f_{T}H^{2}-2f_{\mathcal{T}}\left(
\rho _{m}+p_{m}\right) \right] ,  \label{rhode}
\end{equation}%
\begin{eqnarray}
p_{DE}&=: & \left( \rho _{m}+p_{m}\right)
 \left[ \frac{1+f_{\mathcal{T}}/8\pi G}{1+f_{T}-12H^{2}f_{TT}+H\left( d\rho
_{m}/dH\right) \left( 1-3c_{s}^{2}\right) f_{T\mathcal{T}}}-1\right]
 \nonumber  \\
&&   +\frac{1}{16\pi G}\left[ f+12H^{2}f_{T}-2f_{\mathcal{T}}\left( \rho
_{m}+p_{m}\right) \right] ,
 \label{pde}
\end{eqnarray}%
respectively,  the cosmological field equations of the $f(T,\mathcal{T})$
theory are rewritten in the usual form
\begin{eqnarray}
H^{2} &=&\frac{8\pi G}{3}\left( \rho _{DE}+\rho _{m}\right) ,  \label{Fr2} \\
\dot{H} &=&-4\pi G\left( \rho _{DE}+p_{DE}+\rho _{m}+p_{m}\right) .
\end{eqnarray}
Furthermore, we define the dark energy equation-of-state parameter as
\begin{equation}
w_{DE}=:\frac{p_{DE}}{\rho _{DE}},
\label{wDE1}
\end{equation}
and it proves convenient to introduce also the total equation-of-state parameter $w$, given by
\be
w=:\frac{p_{DE}+p_m}{\rho _{DE}+\rho _m}.
\ee
Note that in the case of the dust universe, with $p_m=0$, we have $w=w_{DE}/\left(1+\rho _m/\rho _{DE}\right)$.

As we can see from Eqs.~(\ref{Fr2}), the matter energy density and pressure, and the effective
dark energy density and pressure, satisfy the
conservation equation
\begin{equation}
\dot{\rho} _{DE}+\dot{\rho }_m+3H\left(\rho _m+\rho
_{DE}+p_m+p_{DE}\right)=0.
\end{equation}
Thus, one obtains an effective interaction between the dark energy and
matter sectors, which is usual in modified matter coupling theories
\cite{Harko:2008qz,Harko:2010mv,Harko:2012hm,Wang:2012rw,Harko:2014gwa,
Harko:2011kv}. Therefore, in the present model the effective dark energy is not conserved alone, and there is an effective coupling between dark energy and  normal matter, with the possibility
of energy transfer from one component to another. The dark energy alone satisfies the ``conservation'' equation
\begin{equation}
\dot{\rho} _{DE}+3H\left(\rho_{DE}+p_{DE}\right)=-Q\left(\rho _m,p_m\right),
\end{equation}
where the effective dark energy ``source'' function $Q \left(\rho _m,p_m\right)$ is 
\be
Q \left(\rho _m,p_m\right)=\dot{\rho} _m+3H\left(\rho _m+p_m\right).
\ee
Hence, in the present model it is allowed to have an energy transfer from ordinary matter to dark energy 
(which, even geometric in its origin, contains a matter contribution), and this process may be interpreted in triggering the accelerating expansion of the universe.

Finally, as an indicator of the accelerating dynamics of the Universe we use
the deceleration parameter $q$, defined as
\begin{equation}
q=-\frac{\dot{H}}{H^2}-1.
\label{deccelpar}
\end{equation}
Positive values of $q$ correspond to decelerating evolution, while negative
values indicates accelerating behavior.

\section{Scalar perturbations and stability analysis}
\label{perturb0}

One of the most important tests in every gravitational theory is the
investigation of the perturbations \cite{Mukhanov:1990me}. Firstly, such a
study reveals the
stability behavior of the theory. Secondly, it allows the correlation of the
gravitational perturbations with the growth of matter overdensities, and
thus one can use growth-index data in order to constrain the parameters of
the scenario. In this section, we examine the scalar perturbations of
$f(T,\mathcal{T})$ gravity at the linear level. Specifically, we extract the
set of gravitational and energy-momentum-tensor perturbations and using
them we examine the stability. Additionally, we extract the equation for the
growth of matter overdensities.

\subsection{Matter and scalar perturbations}
\label{perturb1}

Let us perform a perturbation of the theory. As usual in theories where the
fundamental field is the vierbein, we impose a vierbein perturbation, which
will then lead to the perturbed metric. Without loss of generality we perform
the calculations in the Newtonian gauge.

Denoting the perturbed vierbein with ${e}_{\mu}^A$ and the unperturbed one
with $\bar{e}_{\mu}^A$, the scalar perturbations, keeping up to first-order
terms, write as
\begin{eqnarray}
e_{\mu}^A = \bar{e}_{\mu}^A + t_{\mu}^A,
\end{eqnarray}
with
\begin{eqnarray}
\label{pert1a}
\!\!\!\!
&&\bar{e}_{\mu}^0 = \delta_{\mu}^0,\, \,\,\,\,
\bar{e}_{\mu}^a = \delta_{\mu}^aa,\,\,\,\,\,
\bar{e}^{\mu}_0 = \delta^{\mu}_0, \,\,\,\,\, \bar{e}^{\mu}_a =
\frac{\delta^{\mu}_a}{a},
\\
\!\!\!\!
&&t_{\mu}^0 = \delta_{\mu}^0\psi, \,\,\,\,\, t_{\mu}^a
=-\delta_{\mu}^a a\phi, \,\,\,\,\, t^{\mu}_0 =
-\delta_{0}^{\mu}\psi, \,\,\,\,\, t^{\mu}_a =
\frac{\delta^{\mu}_a}{a}\phi. \ \ \ \ \label{pert2a}
\end{eqnarray}
Note that we have made a simplifying assumption, namely that the scalar
perturbations $t_{\mu}^A$ are diagonal, which is sufficient in order to
study the stability. Furthermore, in this section subscripts zero and one
denote zeroth and linear order values respectively. In the above
expressions we have introduced the scalar modes $\psi$ and
$\phi$, which depend $\bx$ and $t$. The various coefficients have been
considered in a way that the induced metric perturbation to have the usual
form in the Newtonian gauge, that is
\begin{eqnarray}
\label{pertmetric}
 ds^2 = (1 + 2\psi)dt^2
-a^2(1-2\phi)\delta_{ij}dx^idx^j.
\end{eqnarray}

Let us now calculate the various perturbed quantities under the
perturbations
(\ref{pert1a}) and (\ref{pert2a}). Firstly, the vierbein
determinant reads
\begin{eqnarray}
e = \textrm{det}(e_{\mu}^A) = a^3(1+\psi - 3\phi).
\end{eqnarray}
Similarly, the torsion tensor  $T^{\lambda}{}_{\mu\nu}$ and the auxiliary
tensor $S_{\lambda}{}^{\mu\nu}$ read (indices are not summed over):
\begin{eqnarray}
&&T^{0}{}_{\mu\nu} = \partial_{\mu}\psi\delta_{\nu}^0 -
\partial_{\nu}\psi\delta_{\mu}^0,\ \ \ \ T^{i}{}_{0i} = H -
\dot{\phi} \nonumber\\
&&S_{0}{}^{0i} = \frac{\partial_i \phi}{a^2},\ \ \ \
S_{i}{}^{0i} = -H + \dot{\phi} + 2H\psi \nonumber\\
&& T^{i}{}_{ij} = \partial_j\phi ,\ \ \ \  S_{i}{}^{ij} =
\frac{1}{2a^2}\partial_j (\phi - \psi).
\end{eqnarray}
Thus, the torsion scalar can be straightforwardly calculated using
(\ref{torsionscalar}), leading to
\begin{eqnarray}
T =T_0+\delta T,
\end{eqnarray}
where
\begin{eqnarray}
&& T_0=-6H^2\\
&& \delta T=12H(\dot{\phi}+H\psi)
\end{eqnarray}
are respectively the zeroth and first order results.

Having performed the perturbations of the gravitational sector we proceed to
the perturbations of the energy-momentum tensor. As usual they are expressed
as
\begin{eqnarray}
\label{T00pert}
\delta \overset {\mathbf{em}}T_0{}^0 &=& \delta\rho_m\\
\delta \overset {\mathbf{em}}T_0{}^i &=& (\rho_m + p_m)\partial^i
\delta v\\
\delta\overset {\mathbf{em}} T_i{}^0 &=& -a^{2} (\rho_m +
p_m) \partial_i\delta v \label{Ta0pert}
\\
\delta\overset {\mathbf{em}} T_i{}^j &=& -\delta^{j}_{i}\delta
p_m-\partial_i\partial^j\pi^{S}, \label{Tabpert}
\end{eqnarray}
where $\delta{\rho_{m}}$, $\delta{p_{m}}$, $\delta{v}$ are respectively
the fluctuations of energy density, pressure and fluid velocity, while
$\pi^S$ is the scalar component of the anisotropic stress. Additionally,
since $\mathcal{T}\equiv \overset{\mathbf{em}}{T}_{\mu}{}^{\mu}=\overset{\mathbf{em}}{T}_{0}{}^{0}+\overset{\mathbf{em}}{T}_{i}{}^{i}$,  we conclude that
\begin{eqnarray}
\mathcal{T} =\mathcal{T}_0+\delta \mathcal{T},
\end{eqnarray}
where
\begin{eqnarray}
&& \mathcal{T}_0=\rho_{m}-3p_{m}\\
&& \delta \mathcal{T}=\delta\rho_{m}-3\delta p_{m}-\nabla^{2}{\pi^{S}}.
\end{eqnarray} Moreover, we have defined $\nabla^2 =\sum_i \partial_i\partial^i$.

Finally, we can express the variations of the various $f$-derivatives that
appear in the background equations of motion as:
\begin{eqnarray}
&&\delta f=f_T\delta
T+f_{\cal{T}}\delta \cal{T}\nonumber\\
 &&\delta f_T=f_{TT}\delta T+f_{T\cal{T}}\delta \cal{T}\nonumber\\
  &&\delta f_{TT}=f_{TTT}\delta T+f_{TT\cal{T}}\delta \cal{T}\nonumber\\
  &&\delta f_{\cal{T}}=f_{T\cal{T}}\delta T+f_{\cal{T}\cal{T}}\delta
\cal{T}\nonumber\\
  &&\delta f_{T\cal{T}}=f_{TT\cal{T}}\delta
T+f_{T\cal{T}\cal{T}}\delta \cal{T},
\end{eqnarray}
 where the various $f$-derivatives are calculated at the background values
$T_0$ and $\mathcal{T}_0$, for instance  $f_T\equiv
\frac{df}{dT}\Big|_{T=T_0, \mathcal{T}=\mathcal{T}_0}$.

Inserting everything in the equations of motion \eqref{geneoms},  we
acquire the scalar perturbation equations:
\begin{eqnarray}
\label{eqHA}
&&
\left(1+f_{T}\right)\left[\frac{\nabla^{2}{\phi}}{a^{2}}-6H\left(\dot{\phi}
+H\psi\right)\right]
+\left[3H^{2}f_{TT}+\frac{1+f_{T}}{4}-\frac{
\left(\rho_ { m }
+p_{m}\right)f_{T\mathcal{T}}}{2}\right]
\left[12H(\dot{\phi}
+H\psi)\right ]
\notag \\
&&+
\left[3H^{2}f_{T\mathcal{T}}+\frac{f_{\mathcal{T}}}{4}-\frac{\left(\rho_{m}
+p_{m}\right)f_{\mathcal{T}\mathcal{T}}}{2}\right]
\left(\delta{\rho_{
m}}-3 \delta{p_{m}}-\nabla^{2}{\pi^{S}}\right)
\nonumber\\
&&
-\frac{f_{\mathcal{
T } } } { 2 } \left(\delta{\rho_{m}}+\delta{p_{m}}\right)=4\pi G
\,\delta{\rho_{m}},
\end{eqnarray}
\begin{eqnarray}
&&
-\left(1+f_{T}\right)
\partial^{i}{\left(\dot{\phi}+H\psi\right)}
+\left[12H\dot{H}f_{TT}
-\left(\dot{\rho}_{m}-3\dot{p}_{m}\right)f_{T\mathcal{T}}\right]\partial^{i}{
\phi}\nonumber\\
&&
-\frac{a^{2}f_{\mathcal{T}}}{2}\left(\rho_{m}+p_{m}\right)
\partial^{i}{\delta{v}}
=4\pi G a^{2}\left(\rho_{m}+p_{m}\right) \partial^{i}{\delta{v}},
\end{eqnarray}
\begin{eqnarray}
&&
-\left(1+f_{T}\right)
\partial_{i}{\left(\dot{\phi}+H \psi\right)}
+H \partial_{i}\left[12 H f_
{TT} \left(\dot{\phi}+H \psi\right)
+f_{T\mathcal{T}} \left(\delta{\rho_{m}}-3 \delta{p_{m}}-\nabla^{2}{\pi^{S}}
\right)\right]
\notag \\
&&
- \frac{a^{2}
f_{\mathcal{T}}}{2} \left(\rho_{m}+p_{m}\right) \partial_{i}{\delta{v}}
=4 \pi G a^{2} \left(\rho_{m}+p_{m}\right) \partial_{i}{\delta{v}},
\end{eqnarray}
\begin{eqnarray}
&&
\left(1+f_{T}\right) \Big[-H \left(\dot{\psi}+6 \dot{\phi}\right)-2
\psi \left(3 H^{2}+\dot{H}\right)
-\ddot{\phi}+\frac{\nabla^{2}{
\left(\phi-\psi\right)}}{3 a^{2}}\Big]
\notag \\
&&+
 12 H f_{TT} \left[\dot{H} \left(\dot{\phi}+H \psi\right)+H
\left(\ddot{\phi}+\dot{H} \psi+H \dot{\psi}\right)
\right]
+H f_{T\mathcal{T}}
\left(\dot{\delta{\rho}}_{m}-3 \dot{\delta{p}}_{m}-\nabla^{2}{\dot{\pi}^{S}}
\right)
\notag \\
&&
+
\left[12 H \left(\dot{\phi}+H \psi\right)\right]
\left\{f_{TT} \left(3 H^{2}+\dot{H}\right)
-H \left[12 H
\dot{H} f_{TTT
}-f_{T\mathcal{T}T} \left(\dot{\rho}_{m}-3 \dot{p}_{m}\right)\right]+\frac{
1+f_{T}}{4}\right\}
\notag
\\
&&
+ \left(\delta{
\rho_{m}}-3 \delta{p_{m}}-\nabla^{2}{\pi^{S}}\right)
\Big\{f_{T\mathcal{T}} \left(3 H^{2}+\dot{H}\right)
-H \left[12 H \dot{
H} f_{TT\mathcal{T}}-f_{T\mathcal{T}\mathcal{T}} \left(\dot{\rho}_{m}-3
\dot{p}_{m}\right)\right]+\frac{f_{\mathcal{T}}}{4}\Big\}
\notag
\\
&&
+\left(\dot{\phi}+2 H \psi\right) \left[12 H \dot{H} f_{TT}-f_{T \mathcal{T}}
\left(\dot{\rho}_{m}-3
\dot{p}_{m}\right)\right]
+\frac{f_{\mathcal{T}}}{6}
\nabla^{2}{\pi^ {S}}=-4 \pi G \left(
\delta{p}_{m}+\frac{\nabla^{2}{\pi^{S}}}{3}\right),
\label{eq33}
\end{eqnarray}
and
\begin{eqnarray}
\left(1+f_{T}\right)
\left(\psi-\phi\right)=-8 \pi G a^{2} \left(1+\frac{f_{\mathcal{T}}}{8
\pi G}\right) \pi^{S},
\label{phipsi}
\end{eqnarray}
respectively.

\subsection{Stability analysis}
\label{Stab}

Since we have extracted the linear perturbation equations, we can examine
the basic stability requirement by extracting the dispersion relation for
the gravitational perturbations. As usual, for simplicity we will consider
zero anisotropic stress ($\pi^{S}=0$), and in this case equation
(\ref{phipsi}) allows us to replace $\psi$ by $\phi$, and thus remaining
with only one gravitational perturbative degree of freedom. We transform it
in the Fourier space as
\begin{eqnarray}
 \phi(t,\bx)=\int \frac{d^3k}{(2\pi)^\frac{3}{2}}
~\tilde{\phi}_k(t)e^{i\bk\cdot\bx},
\label{phiexpansion}
\end{eqnarray}
and therefore $\nabla^2\phi=-k^2 \tilde{\phi}_k$.

Inserting this decomposition into (\ref{eq33}), and using the other
perturbative equations in order to eliminate variables, after some algebra
we obtain the following equation of motion for the modes of
the gravitational potential $\phi$:
\begin{eqnarray}
\label{phiddk}
\ddot{\tilde{\phi}}_k+\Gamma \dot{\tilde{\phi}}_k+\mu^2
\tilde{\phi}_k+c_s^2\frac{k^2}{a^2}
\tilde{\phi}_k=D.
\end{eqnarray}
The functions $\Gamma$, $\mu^2$ and $c_s^2$ are
respectively the frictional
term, the effective mass, and the sound speed parameter for the
gravitational potential $\phi$, and along with the term $D$ are given in the
Appendix. Clearly, in order for our theory to be stable at the linear scalar
perturbation level, we require  $\mu^2\geq0$ and $c_s^2\geq0$.

Due to the complexity of the coefficients $\mu^2$ and $c_s^2$, we cannot
extract analytical relations for the stability conditions. This is usual in
complicated modified gravity models, for instance in generalized Galileon
theory \cite{DeFelice:2010pv,DeFelice:2011bh}, in Ho\v{r}ava-Lifshitz gravity
\cite{Bogdanos:2009uj,Wang:2009yz}, in cosmology with non-minimal
derivative coupling \cite{Dent:2013awa}, etc. Furthermore, although in
almost all modified gravity models one can, at first stage, perform the
perturbations neglecting the matter sector, in the scenario at hand this
cannot be done, and this is an additional complexity, since in that case one
would kill the extra information of the model (which comes from the matter
sector itself) remaining with the usual $f(T)$ gravity. A significant
simplification arises if we consider as usual the matter to be dust, that is
$p_m=\delta p_m=0$, but still one needs to resort to numerical elaboration
of equation (\ref{phiddk}) in order to ensure if a given $f(T,\mathcal{T})$
cosmological model is free of instabilities. However, we mention that
since the simple $f(T)$ gravity is free of instabilities for a large
class of $f(T)$ ansatzes \cite{Chen:2010vaoooo,Dent:2011zz}, we deduce that at least for
$f(T,\mathcal{T})$ models that are small deviations from the corresponding
$f(T)$ ones, the stability requirements $\mu^2\geq0$ and $c_s^2\geq0$ are
expected to be satisfied.

\section{Cosmological behavior}
\label{Cosmologicalbehavior}

In this section, we investigate the cosmological implications
of   $f(T,\mathcal{T})$ gravity, focusing on specific examples. For
convenience, we use the natural system of units with $8\pi G=c=1$. From the
analysis of the previous section we saw that the basic equations
describing the cosmological dynamics are the two Friedmann equations
(\ref{Friedmann1}) and (\ref{F2}). These can be re-written as
\begin{equation}
\rho _{m}=\frac{3H^{2}+\left( f+12H^{2}f_{T}|_{T\rightarrow -6H^{2}}\right)
/2-f_{\mathcal{T}}p_{m}}{1+f_{\mathcal{T}}},  \label{eq1}
\end{equation}%
and
\begin{equation}
\dot{H}= -\frac{\left( 1+f_{\mathcal{T}}\right) \left( \rho
_{m}+p_{m}\right) /2+H\left( \dot{\rho}_{m}-3\,\dot{p}_{m}\right) f_{T%
\mathcal{T}}|_{T\rightarrow -6H^{2}}}{1+f_{T}|_{T\rightarrow
-6H^{2}}-12H^{2}f_{TT}|_{T\rightarrow -6H^{2}}},  \label{eq2}
\end{equation}
respectively. Equations (\ref{eq1}) and (\ref{eq2}) compose a system of two
differential equations for three unknown functions, namely $\left(H,\rho
_{m},p_{m}\right) $. In order to close the system of equations we need to
impose the matter equation of state   $p_{m}=p_{m}\left(\rho
_{m}\right)$. In this work, we restrict our study to the case of dust
matter, that is $p_{m}=0$, and thus $\mathcal{T}=\rho _{m}$.

In the following, we investigate two specific  $f(T,\mathcal{T})$ models,
corresponding to simple non-trivial extensions of TEGR, that is of GR.
However, although simple, these models  reveal the new features and the
capabilities of the theory.

In order to relate our model with cosmological observations we will present the results of the numerical computations for the Hubble function, matter energy density, deceleration parameter and the parameter of the dark energy equation of state as functions of the cosmological redshift $z$, defined as
\begin{equation}
z=\frac{a_0}{a}-1,
\end{equation}
where $a_0$ is the present day value of the scale factor, which we take as one, that is, we choose $a_0=1$. In terms of the redshift the derivatives with respect to time are expressed as
\begin{equation}
\frac{d}{dt}=-(1+z)H(z)\frac{d}{dz}.
\end{equation}
In particular for the deceleration parameter we obtain
\begin{equation}
q(z)=\frac{1+z}{H(z)}\frac{dH}{dz}-1.
\end{equation}

In order to numerically integrate the gravitational field equations we need to fix the value of the Hubble function at $z=0$, $H(0)=H_0$. The present value of the Hubble function is of the order of $H_0\approx 2.3\times 10^{-18}$ s$^{-1}$ \cite{Planck}.

\subsection{Model A: $f\left(T,\mathcal{T}\right)=\protect\alpha \protect
T^n\,\mathcal{T}+\Lambda$}

A first model describing a simple departure from General Relativity is
the one with  $f\left(T,\mathcal{T}\right)= \alpha
T^n \mathcal{T}+\Lambda=\alpha T^n \rho _m+\Lambda$, where $\alpha $, $n\neq
0$ and $\Lambda $ are arbitrary constants.
For this ansatz, we straightforwardly obtain
%
$f=\alpha \left(-6H^2\right)^n\rho _m +\Lambda$,    $f_T=n\alpha
\rho_m\left(-6H^2\right)^{n-1}$, $f_{TT}=\alpha
n(n-1)\left(-6H^2\right)^{n-2}$, $f_{T\mathcal{T}}=\alpha n \left(-6H^2\right)^{n-1}$, and $f_{\mathcal{T}}=\alpha \left(-6 H^2\right)^n$.
Hence, inserting these into Eq.~(\ref{eq1}) we can obtain the matter energy density as a function of the
Hubble function as
\begin{equation}
\label{eqrA}
\rho _m=\frac{3H^2+\Lambda /2}{1+\alpha (n+1/2)\left(-6 H^2\right)^n}.
\end{equation}
Differentiating Eq.~(\ref{eqrA}) we acquire the useful relation
\begin{equation}
\label{eq11}
\dot{\rho}_m=\frac{2 \dot{H} \left[12 H^2-(2 n+1)\alpha  6^n
\left(-H^2\right)^n \left(6
   (n-1) H^2+\Lambda  n\right)\right]}{H \left[(2 n+1)\alpha  6^n
   \left(-H^2\right)^n+2\right]^2}.
\end{equation}
Thus, inserting the above expressions into Eqs. (\ref{wDE1}),
(\ref{deccelpar}) and (\ref{eq2}) we extract respectively the
time-variation of the Hubble function, the deceleration parameter and the
dark-energy equation-of-state parameter, as functions of $H$, namely
\begin{eqnarray}
&\label{eqHAb}
\dot{H}=-\frac{3H^{2}\left( 6H^{2}+\Lambda \right) \left[ \alpha 6^{n}\left(
-H^{2}\right) ^{n}+1\right] \left[ \alpha 6^{n}(2n+1)\left( -H^{2}\right)
^{n}+2\right] }{\alpha ^{2}36^{n}(2n+1)\left( -H^{2}\right) ^{2n}\left[
6(n+1)H^{2}+\Lambda n\right] -\alpha 2^{n+1}3^{n}\left( -H^{2}\right) ^{n}%
\left[ 6(n-2)(2n+1)H^{2}+\Lambda n(2n-1)\right] +24H^{2}},
\end{eqnarray}
\begin{eqnarray}\label{eqqA}
&q=\frac{3\left( 6H^{2}+\Lambda \right) \left[ \alpha 6^{n}\left(
-H^{2}\right) ^{n}+1\right] \left[ \alpha 6^{n}(2n+1)\left( -H^{2}\right)
^{n}+2\right] }{\alpha ^{2}36^{n}(2n+1)\left( -H^{2}\right) ^{2n}\left[
6(n+1)H^{2}+\Lambda n\right] -\alpha 2^{n+1}3^{n}\left( -H^{2}\right) ^{n}%
\left[ 6(n-2)(2n+1)H^{2}+\Lambda n(2n-1)\right] +24H^{2}}-1,
\end{eqnarray}
and
\begin{eqnarray}
\label{wn}
&w_{DE}=-
\frac{3H^{2}\left[ \alpha 6^{n}(2n+1)\left( -H^{2}\right) ^{n}+2\right]
\left\{ \alpha _{1}\alpha _{3}\left( -H^{2}\right) ^{n}H^{2}+\alpha
_{4}-\alpha _{2}\left( -H^{2}\right) ^{2n}\left[ 6(n-1)H^{2}+\Lambda (n-2)%
\right] +4\Lambda \right\} }{\left[ \alpha _{1}(2n+1)\left( -H^{2}\right)
^{n+1}+\Lambda \right] \left\{ \alpha _{2}\left( -H^{2}\right) ^{2n}\left[
6(n+1)H^{2}+\Lambda n\right] -\alpha _{1}\left( -H^{2}\right) ^{n}\left[
\alpha _{5}H^{2}+\alpha _{6}\right] +24H^{2}\right\} },
\end{eqnarray}
respectively, where for convenience we have defined the parameters $\alpha _{1}=\alpha
2^{n+1}3^{n}$, $\alpha _{2}=\alpha
^{2}36^{n}\left( 2n+1\right) $, $\alpha_{3}=6[n(2n-1)+1]$, $\alpha
_{4}=\Lambda \left( 2n^{2}+n+3\right) $, $\alpha _{5}=6(n-2)(2n+1)$, and $
\alpha _{6}=\Lambda n(2n-1)$.

\subsubsection{The case $n=1$}

A first model describing the simplest departure from General Relativity is
the one obtained for $n=1$ in the general scenario previously introduced,
that is with $f\left(T,\mathcal{T}\right)= \alpha
T \mathcal{T}=\alpha T \rho _m+\Lambda$.
For this ansatz, we straightforwardly obtain  $f=-6\alpha \rho
_m H^2+\Lambda $,  $f_T=\alpha \rho _m$, $f_{TT}=0$, $f_{T%
\mathcal{T}}=\alpha $, and $f_{\mathcal{T}}=\alpha T=-6\alpha H^2$. Thus,
Eq.~(\ref{eqrA}) reduces to
\be\label{eqrA1}
\rho _m=\frac{3 H^2+\Lambda /2}{1-9 \alpha  H^2},
\ee
while from Eqs. (\ref{eqHAb})--(\ref{wn}) we obtain
\be
\label{eqHA1}
\dot{H}= -\frac{\left(6 \alpha  H^2-1\right) \left(9 \alpha  H^2-1\right)
\left(6 H^2+\Lambda \right)}{2 \left[\alpha  \Lambda +9 \alpha  H^2
   \left(\alpha  \Lambda +12 \alpha  H^2-2\right)+2\right]},
\ee
\be
\label{eqqA1}
q=\frac{\left(6 \alpha  H^2-1\right) \left(9 \alpha  H^2-1\right) \left(6
H^2+\Lambda \right)}{2 H^2 \left[\alpha  \Lambda +9 \alpha  H^2 \left(\alpha
   \Lambda +12 \alpha  H^2-2\right)+2\right]}-1,
\ee
and
\begin{equation}
\label{eqwA}
w_{DE}=\frac{2 \left(9 \alpha  H^2-1\right) \left[9 \alpha  (3
\alpha
\Lambda -4) H^4-18 \alpha  \Lambda  H^2+\Lambda \right]}{\left(54 \alpha
H^4+\Lambda \right)
   \left[\alpha  \Lambda +9 \alpha  H^2 \left(\alpha  \Lambda +12 \alpha
H^2-2\right)+2\right]},
\end{equation}
respectively.
Note that relations (\ref{eqrA1})--(\ref{eqqA1}) hold for every
$\alpha$, including $\alpha=0$ (in which case we obtain the GR expressions),
while  (\ref{eqwA}) holds for $\alpha\neq0$, since for $\alpha=0$ the
effective dark energy sector does not exist at all (both $\rho_{DE}$ and
$p_{DE}$ are zero).

As we may observe from Eq.~(\ref{eqqA1}), the scenario at hand can give rise to both
acceleration and deceleration phases, according to the values of the model
parameters
$\alpha$ and $\Lambda $. However, the most interesting feature that is
clear from Eq. (\ref{eqwA}) is that the dark energy
equation-of-state parameter can be quintessence-like or phantom-like, or
even experience the phantom-divide crossing during the evolution,
depending on the choice of the parameter range.
This feature is an additional advantage, since such behaviors are difficult to be
obtained in dark energy constructions.

\begin{figure}[ht]
\centering
\includegraphics[width=8cm]{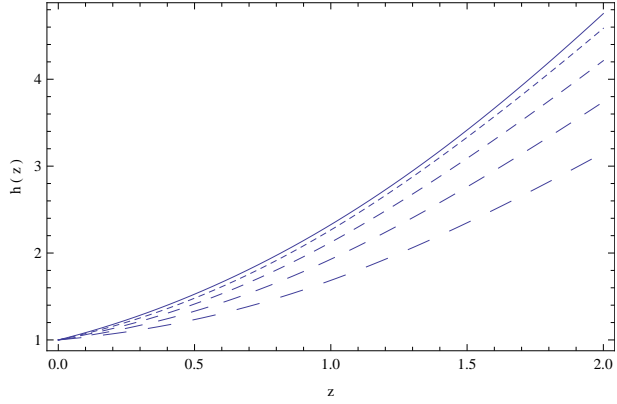}
\caption{Variation of the dimensionless Hubble function $h(z)$ as a function of the redshift $z$ for the model
$f(T,\mathcal{T})=\protect\alpha \protect\mathcal{T}T+\Lambda $, for five
different choices of the parameters $\protect\alpha _0$, and $\lambda$:
  $\protect\alpha _0=-0.01$, $\lambda =-3$ (solid curve), $\protect\alpha _0
=-0.02$, $\lambda =-3.5$ (dotted curve), $%
\protect\alpha _0=-0.03$, $\lambda =-4$ (short-dashed curve), $\protect\alpha _0
=-0.04$, $\lambda =-4.5$ (dashed
curve), and $\protect\alpha _0=-0.05$, $\lambda =-5$ (long-dashed curve),
respectively. }
\label{fig1}
\end{figure}

\begin{figure}[!]
\centering
\includegraphics[width=8cm]{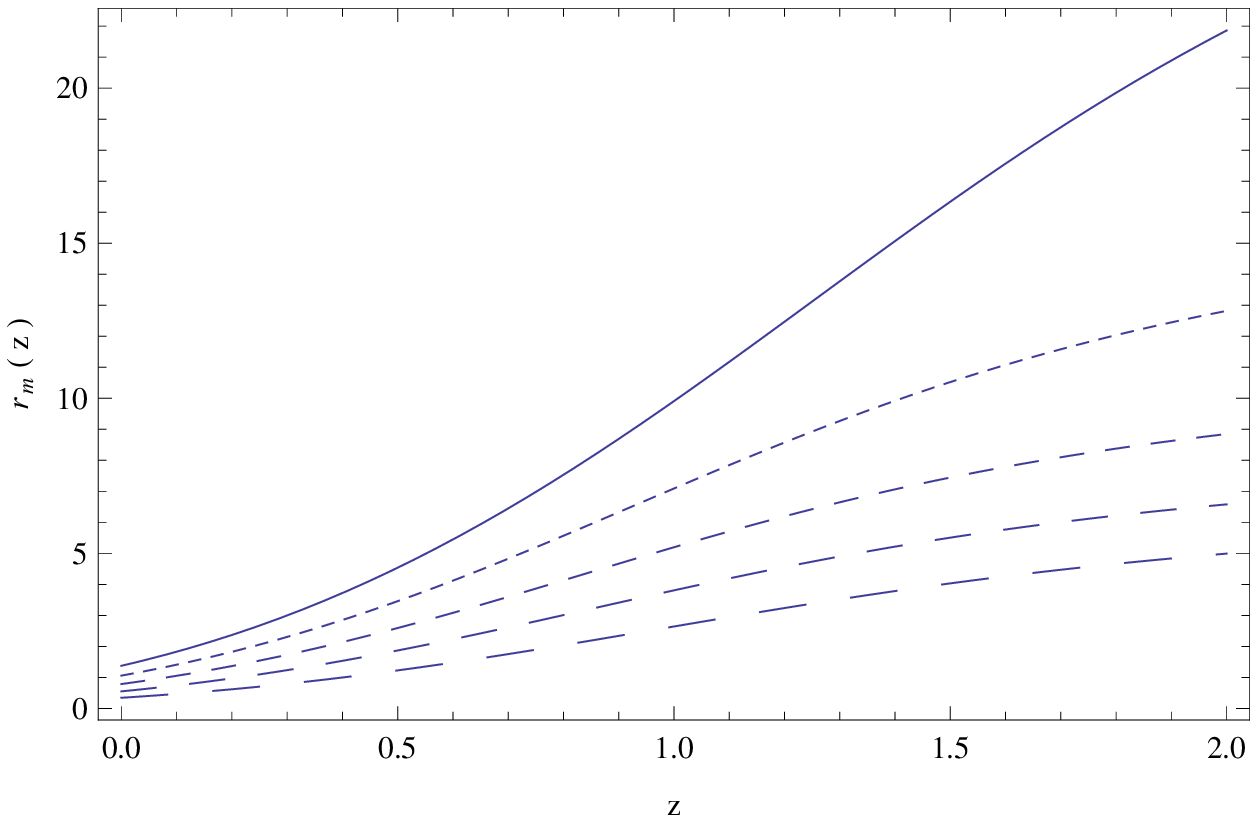}
\caption{Variation of the dimensionless matter energy density $r _m(z)$ as a function of the redshift $z$
 for the model
$f(T,\mathcal{T})=\protect\alpha \protect\mathcal{T}T+\Lambda $, for five
different choices of the parameters $\protect\alpha _0$, and $\lambda$:
  $\protect\alpha _0=-0.01$, $\lambda =-3$ (solid curve), $\protect\alpha _0
=-0.02$, $\lambda =-3.5$ (dotted curve), $%
\protect\alpha _0=-0.03$, $\lambda =-4$ (short-dashed curve), $\protect\alpha _0
=-0.04$, $\lambda =-4.5$ (dashed
curve), and $\protect\alpha _0=-0.05$, $\lambda =-5$ (long-dashed curve),
respectively.}
\label{fig2}
\end{figure}

\begin{figure}[!]
\centering
\includegraphics[width=8cm]{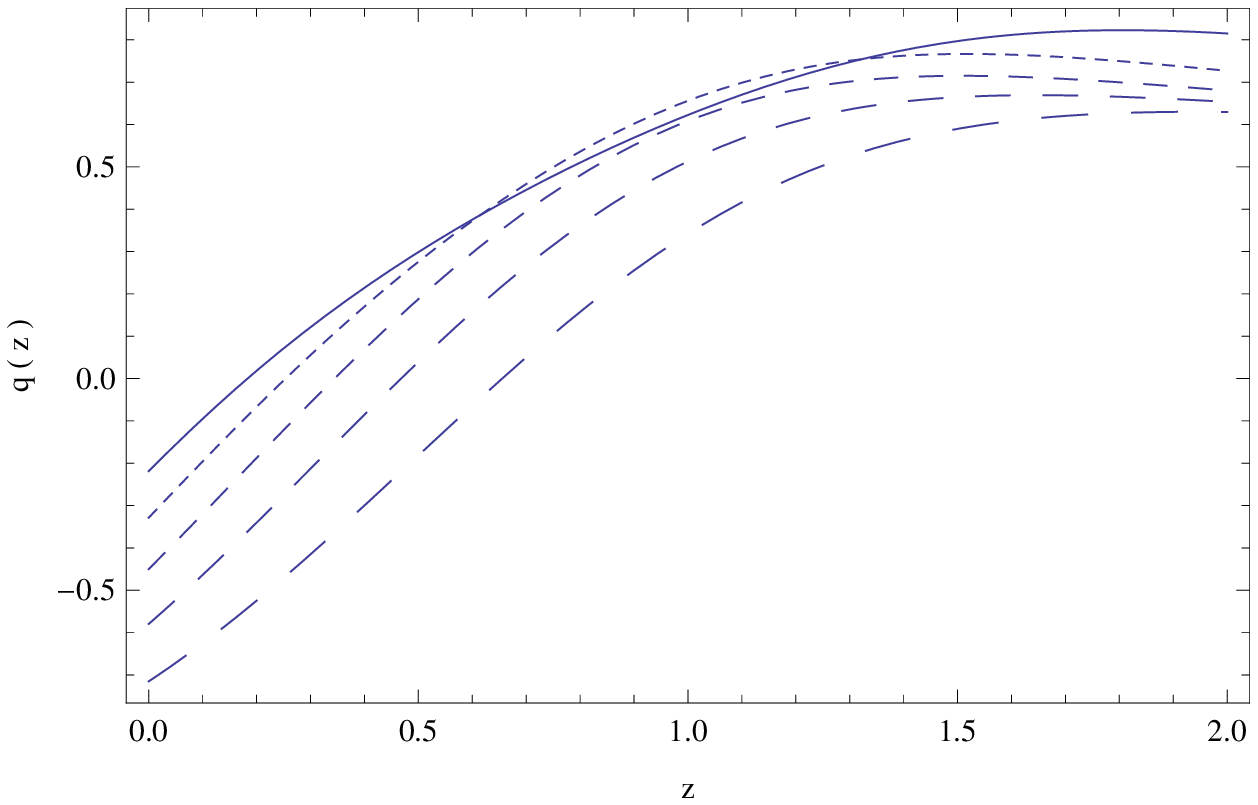}
\caption{Variation of the deceleration parameter $q(z)$  as a function of the redshift $z$ for the model
$f(T,\mathcal{T})=\protect\alpha \protect\mathcal{T}T+\Lambda $, for five
different choices of the parameters $\protect\alpha _0$, and $\lambda$:
  $\protect\alpha _0=-0.01$, $\lambda =-3$ (solid curve), $\protect\alpha _0
=-0.02$, $\lambda =-3.5$ (dotted curve), $%
\protect\alpha _0=-0.03$, $\lambda =-4$ (short-dashed curve), $\protect\alpha _0
=-0.04$, $\lambda =-4.5$ (dashed
curve), and $\protect\alpha _0=-0.05$, $\lambda =-5$ (long-dashed curve),
respectively.}
\label{fig3}
\end{figure}

\begin{figure}[!]
\centering
\includegraphics[width=8cm]{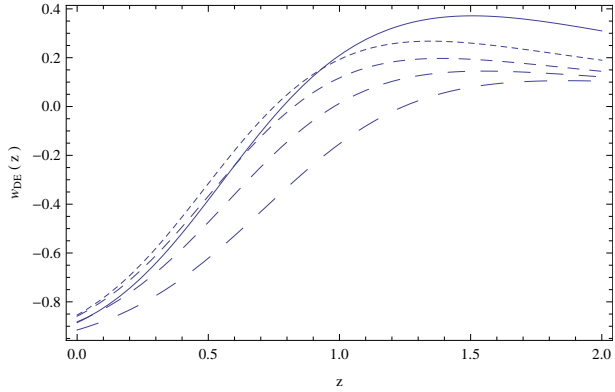}
\caption{Variation of the parameter of the dark energy equation of state
$w_{DE}(z)$ as a function of $z$ for the model
$f(T,\mathcal{T})=\protect\alpha \protect\mathcal{T}T+\Lambda $, for five
different choices of the parameters $\protect\alpha _0$, and $\lambda$:
  $\protect\alpha _0=-0.01$, $\lambda =-3$ (solid curve), $\protect\alpha _0
=-0.02$, $\lambda =-3.5$ (dotted curve), $%
\protect\alpha _0=-0.03$, $\lambda =-4$ (short-dashed curve), $\protect\alpha _0
=-0.04$, $\lambda =-4.5$ (dashed
curve), and $\protect\alpha _0=-0.05$, $\lambda =-5$ (long-dashed curve),
respectively.}
\label{fig4}
\end{figure}

\begin{figure}[!]
\centering
\includegraphics[width=8cm]{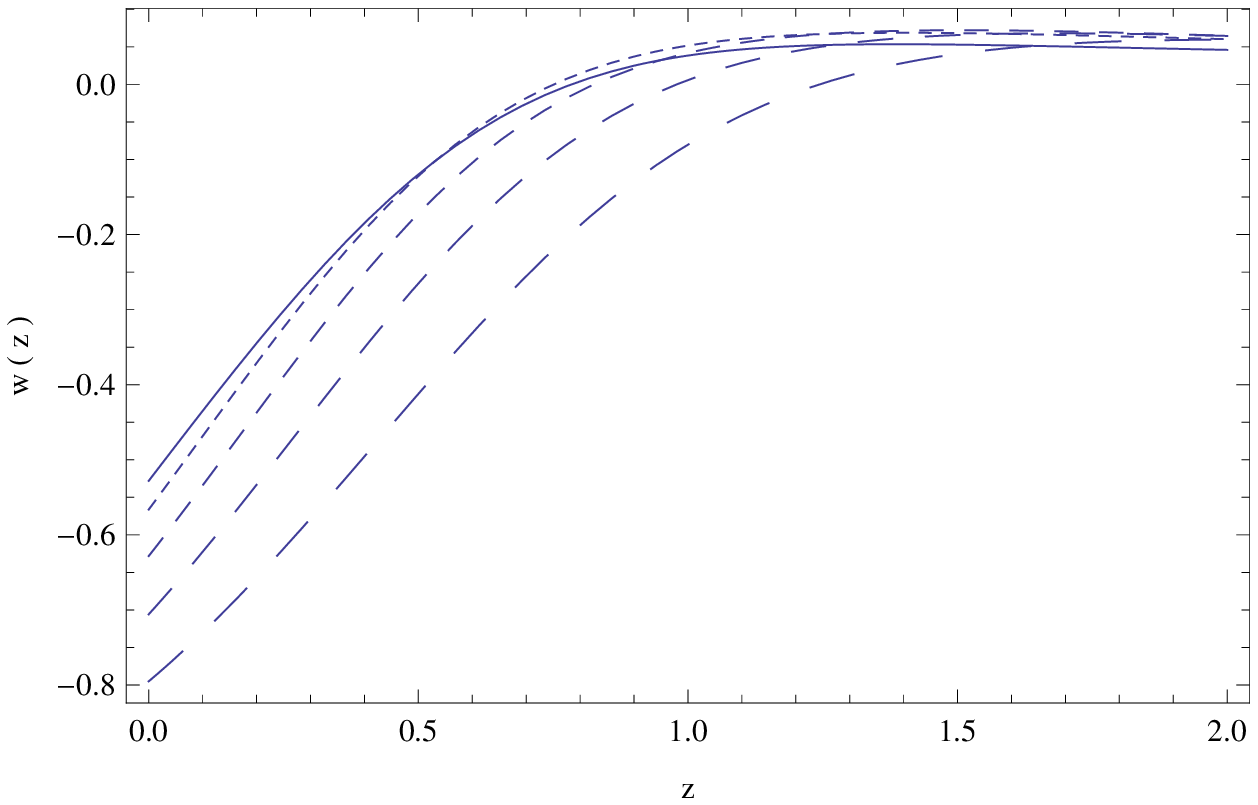}
\caption{Variation of the total equation-of-state parameter
$w$ as a function of $z$ for the model
$f(T,\mathcal{T})=\protect\alpha \protect\mathcal{T}T+\Lambda $, for five
different choices of the parameters $\protect\alpha _0$, and $\lambda$:
  $\protect\alpha _0=-0.01$, $\lambda =-3$ (solid curve), $\protect\alpha _0
=-0.02$, $\lambda =-3.5$ (dotted curve), $%
\protect\alpha _0=-0.03$, $\lambda =-4$ (short-dashed curve), $\protect\alpha _0
=-0.04$, $\lambda =-4.5$ (dashed
curve), and $\protect\alpha _0=-0.05$, $\lambda =-5$ (long-dashed curve),
respectively.}
\label{fig4n}
\end{figure}

In order to present the above features in a more transparent way, we proceed
to a detailed numerical elaboration for various parameter choices. We introduce the redshift $z$ as the independent variables, and we rescale the parameters as
\begin{equation}
H(z)=H_0h(z), \qquad  \rho _m(z)=r_m(z)H_0^2, \qquad \Lambda =\lambda H_0^2, \qquad  \alpha =\frac{\alpha _0}{H_0^2},
\end{equation}
where $H_0$ is the present value of the Hubble function, and $\left(r_m,\lambda ,\alpha _0\right)$ represent the dimensionless matter density, and the dimensionless model parameters. Therefore Eqs.~(\ref{eqrA1}) - (\ref{eqwA}) take the form
\begin{equation}
r_m=\frac{3h^2+\lambda /2}{1-9\alpha _0h^2},
\end{equation}
\bea\label{zA}
(1+z)h\frac{dh}{dz}=\frac{\left(6 \alpha _0  h^2-1\right) \left(9 \alpha_0  h^2-1\right)
\left(6 h^2+\lambda \right)}{2 \left[\alpha _0 \lambda +9 \alpha _0  h^2
   \left(\alpha _0  \lambda +12 \alpha_0  h^2-2\right)+2\right]},\nonumber\\
   \eea
\be
q=\frac{\left(6 \alpha _0  h^2-1\right) \left(9 \alpha _0 h^2-1\right) \left(6
h^2+\lambda \right)}{2 h^2 \left[\alpha _0 \lambda +9 \alpha _0  h^2 \left(\alpha _0
   \lambda +12 \alpha _0  h^2-2\right)+2\right]}-1,
\ee
\be
w_{DE}=\frac{2 \left(9 \alpha_0  h^2-1\right) \left[9 \alpha_0  (3
\alpha _0
\lambda -4) h^4-18 \alpha _0  \lambda  h^2+\lambda \right]}{\left(54 \alpha _0
h^4+\lambda \right)
   \left[\alpha _0  \lambda +9 \alpha _0  h^2 \left(\alpha  \lambda +12 \alpha _0
h^2-2\right)+2\right]}.
\ee
Eq.~(\ref{zA}) must be integrated with the initial condition $h(0)=1$.
In
Figs.~\ref{fig1}-\ref{fig4n}, we depict the corresponding results, namely the
redshift-variation of the Hubble function,  of
the matter energy density, of the deceleration parameter, of the
parameter of the dark energy equation of state, and of the total equation of state, respectively. We mention
that for all these evolutions, we have numerically verified that the
stability conditions extracted in section \ref{perturb0} are satisfied.

As one can see from the Figures, depending on the values of the parameters
$\alpha$ and $\Lambda$, the Universe can exhibit a very interesting
dynamics. The Hubble function, presented in Fig.~\ref{fig1}, is a
monotonically decreasing function of time (monotonically increasing function of the redshift) during the entire evolution of the
considered redshift range of the Universe. The scale factor is an increasing function of time, and  the matter
energy density, plotted in Fig.~\ref{fig2}, tends to zero in the large-time
limit. As one can see from Fig.~\ref{fig3}, the dust filled Universe starts its evolution at the redshift $z=2$
from a decelerating state, with $q\approx 0.5-0.8>0$. At around $z\approx 0.5$, $q\approx 0$, and the Universe enters in an accelerating phase, with $q$ tending towards $-1$ at around $z=0$.  This
evolution is in agreement with the observed behavior of the recent Universe,
namely a first decelerating matter dominated stage, a transition to accelerating expansion, and then the transition to late-time accelerating
phase. Note that at asymptotically large times the Universe ends
in a de Sitter expansion.

The parameter $w_{DE}$ of the dark energy equation of state, presented in Fig.~\ref{fig4}, shows a similar evolution, tending towards minus one at $z=0$, when the Universe enters in a de Sitter phase, with its dynamics dominated by the effective dark energy component, mimicking a cosmological constant. Additionally, in Fig.~\ref{fig4n} we
present the   total equation-of-state parameter $w=w_{DE}/\left(1+\rho _m/\rho _{DE}\right)$, and we can observe  a dynamics similar to $w_{DE}$.
Finally, for these specific parameter choices both the dark energy
equation-of-state parameter, as well as the total one,  lie in the quintessence
regime, approaching the cosmological constant value $-1$ at large times (as $\rho_{DE}$ becomes larger and 
larger comparing to $\rho_m$, $w$ tends to coincide with $w_{DE}$).

We close this analysis by examining the limiting behavior of the model.
In the limit $\alpha H^2 \ll 1$ and $\alpha \Lambda
\ll 1$,   Eqs.~(\ref{eqrA1}) and (\ref{eqHA1}) become
\begin{eqnarray}
\rho _m&=&3H^2+\frac{\Lambda
}{2} \\
 \dot{H}&=&-\frac{3}{2}H^2+\frac{\Lambda }{4}.
\end{eqnarray}
The above relationships, in the large-time limit and for $\Lambda < 0$, provide the standard de Sitter
cosmological evolution, with $q=-1$, $H=H_0=\sqrt{\Lambda /6}$ and $a\propto
\exp\left(H_0t\right)$. Note that this limit is valid independently of the
$\alpha$-value. However, for $\alpha >0$  the positivity of the matter energy
density constraints the $\alpha$-values in the region that leads to
$9\alpha H^2<1$.

On the other hand, for $ \alpha H^2 \gg 1$ the matter energy density tends to
\be
\rho _m=\frac{1}{3\alpha }+\frac{\Lambda }{18\alpha H^2},
\ee
while the dynamics of the Hubble function is determined by the equation
\be
\dot{H}=-\frac{3}{2}H^2+\frac{\Lambda }{4}.
\ee
Thus, the general solution given by
\be
H(t)= \sqrt{\frac{\Lambda}{6}}\;  \tanh \left[\frac{\sqrt{6\Lambda }}{4} \left(  t-4
    C_1 \right)\right],
\ee
where $C_1$ is an arbitrary constant of integration.

\subsubsection{ The case  $n\neq 1$}

\begin{figure}[ht]
\centering
\includegraphics[width=8cm]{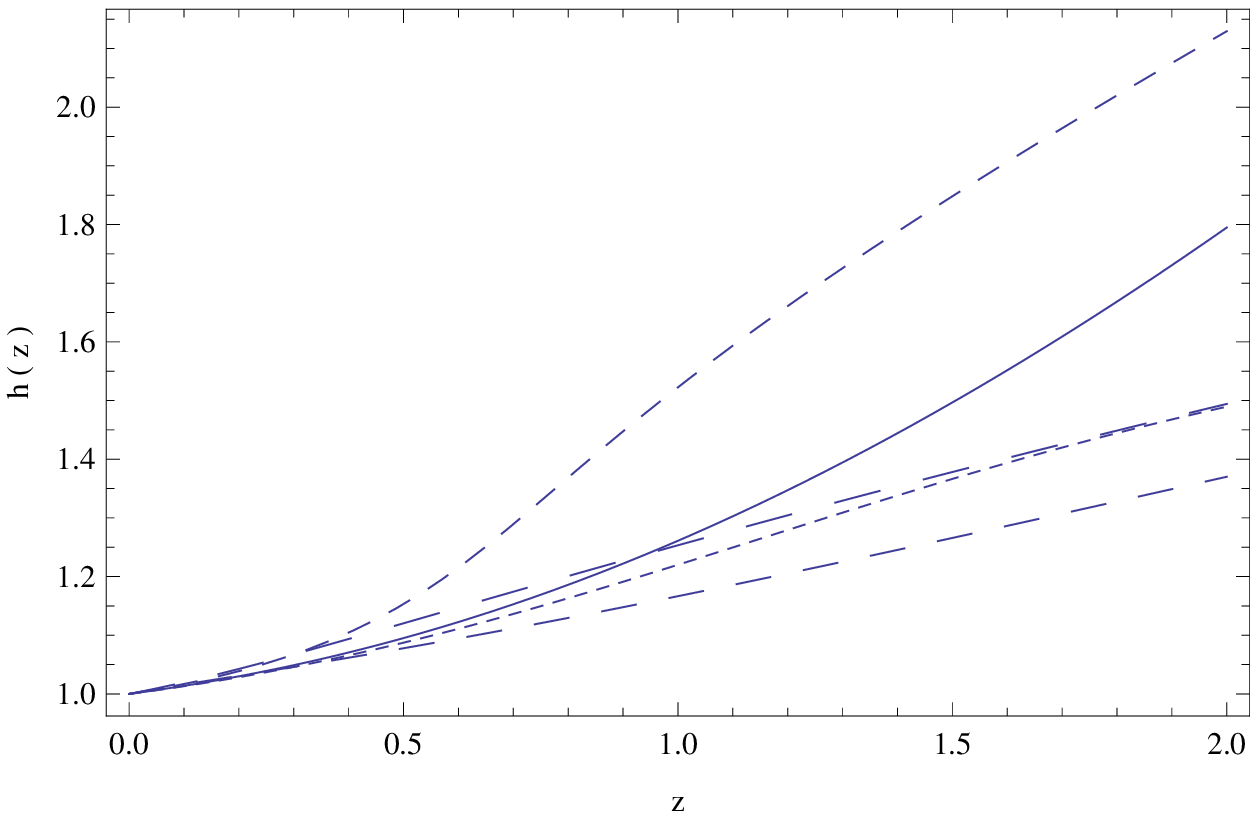}
\caption{Variation of the dimensionless Hubble function $h(z)$ as a function of the redshift $z$ in the $f(T,\mathcal{T%
})$ gravity theory with $f(T,\mathcal{T})=\protect\alpha \protect\rho
_mT^n+\Lambda $,
for $\alpha _0=-0.0011$, $\lambda =-5.5$, and for five different values of
$n$:
$n=1$ (solid curve), $n=2$ (dotted curve),
 $n=3$ (short-dashed curve),  $n=4$ (dashed
curve), and $n=5$, respectively. }
\label{fig1a}
\end{figure}

\begin{figure}[!]
\centering
\includegraphics[width=8cm]{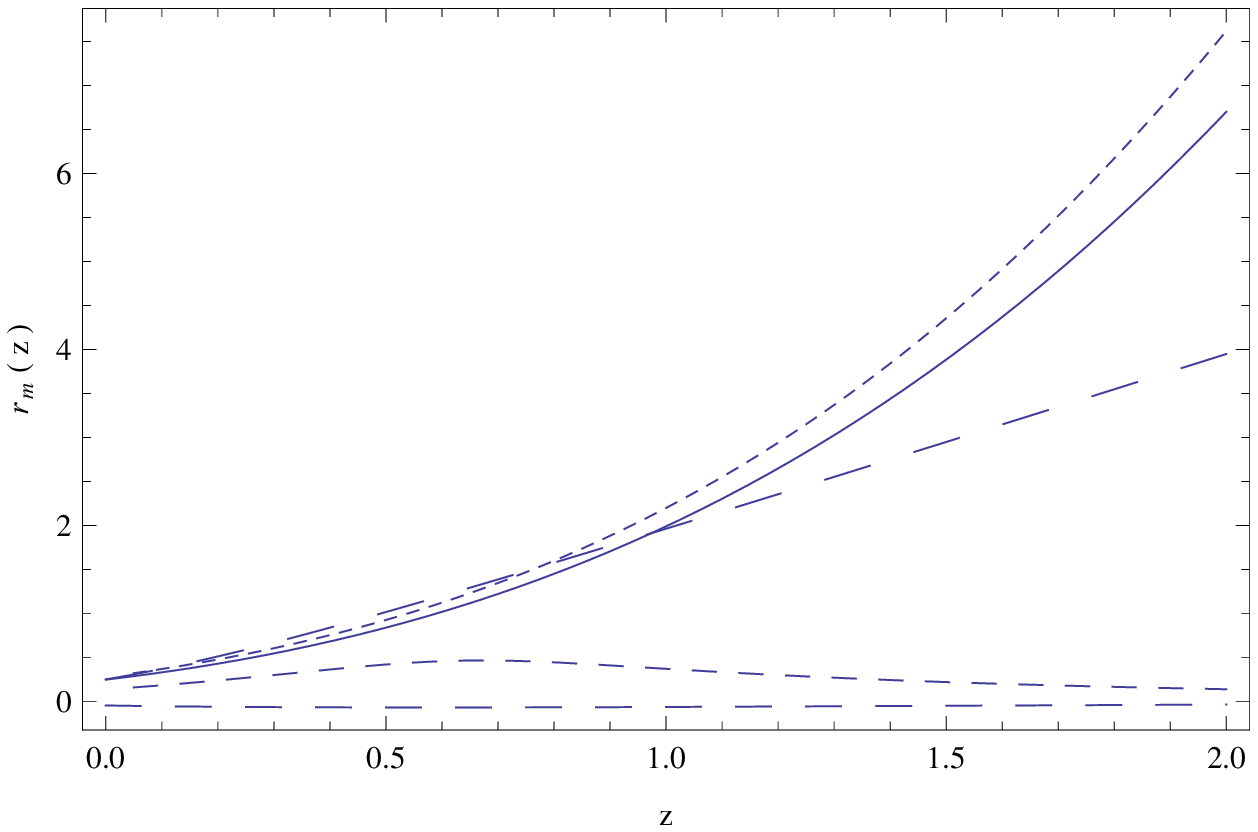}
\caption{Variation of the dimensionless matter energy density $\rho _m(z)$ as a function of the redshift $z$
in the $f(T,\mathcal{T%
})$ gravity theory with $f(T,\mathcal{T})=\protect\alpha \protect\rho
_mT^n+\Lambda $,
for $\alpha _0=-0.0011$, $\lambda =-5.5$, and for five different values of
$n$:
$n=1$ (solid curve), $n=2$ (dotted curve),
 $n=3$ (short-dashed curve),  $n=4$ (dashed
curve), and $n=5$, respectively.}
\label{fig3a}
\end{figure}

\begin{figure}[!]
\centering
\includegraphics[width=8cm]{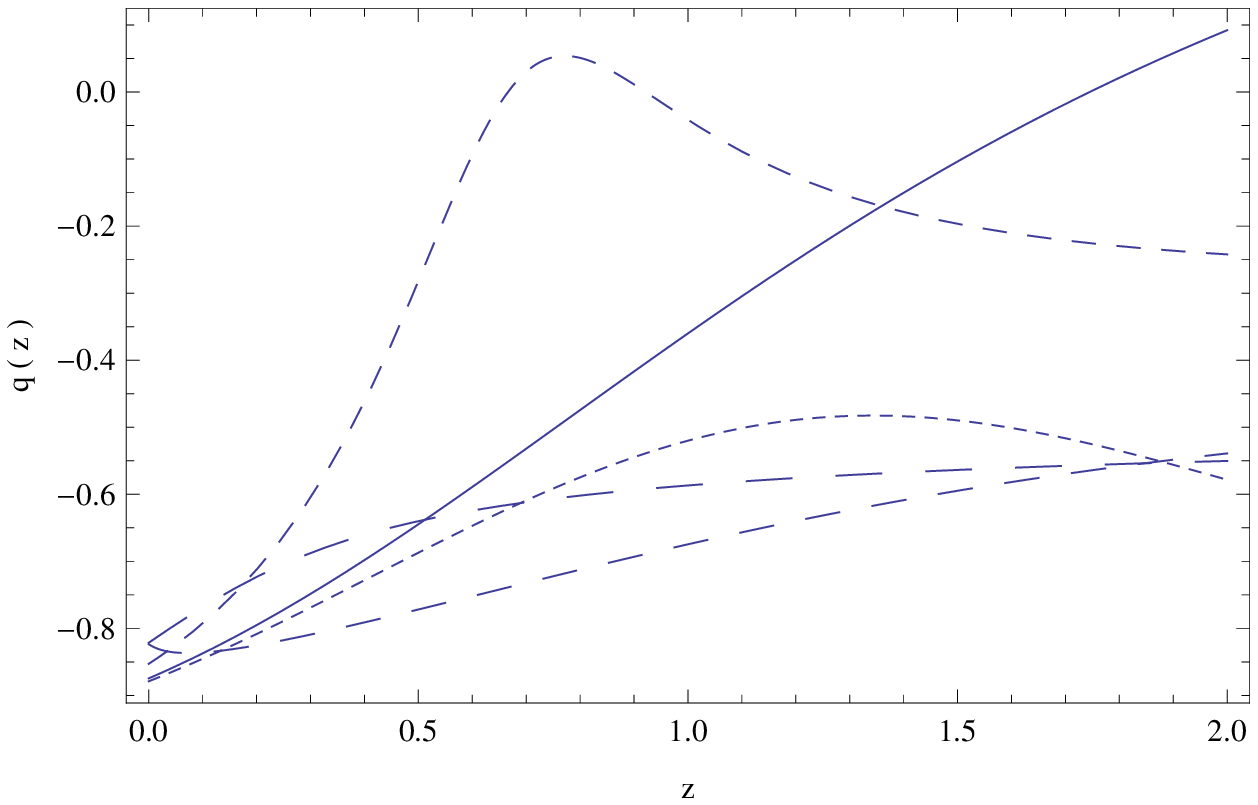}
\caption{Variation of the deceleration parameter $q(z)$ as a function of the redshift $z$ in the
$f(T,\mathcal{T%
})$ gravity theory with $f(T,\mathcal{T})=\protect\alpha \protect\rho
_mT^n+\Lambda $,
for $\alpha _0=-0.0011$, $\lambda =-5.5$, and for five different values of
$n$:
$n=1$ (solid curve), $n=2$ (dotted curve),
 $n=3$ (short-dashed curve),  $n=4$ (dashed
curve), and $n=5$, respectively.}
\label{fig3a2}
\end{figure}

\begin{figure}[!]
\centering
\includegraphics[width=8cm]{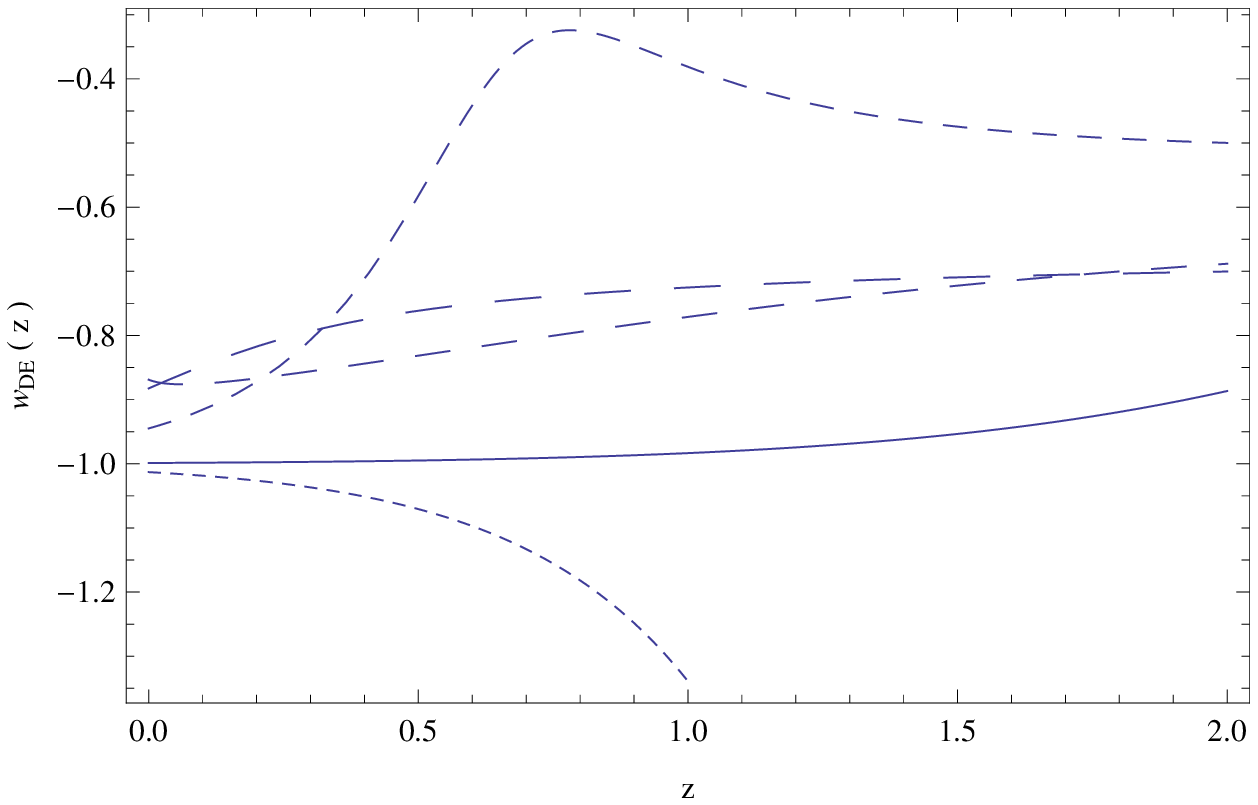}
\caption{Variation of the parameter of the dark energy equation of state
$w_{DE}(z)$ as a function of $z$ in  the $f(T,\mathcal{T%
})$ gravity theory with $f(T,\mathcal{T})=\protect\alpha \protect\rho
_mT^n+\Lambda $,
for $\alpha _0=-0.0011$, $\lambda =-5.5$, and for five different values of
$n$:
$n=1$ (solid curve), $n=2$ (dotted curve),
 $n=3$ (short-dashed curve),  $n=4$ (dashed
curve), and $n=5$, respectively.}
\label{fig4a}
\end{figure}

\begin{figure}[ht]
\centering
\includegraphics[width=8cm]{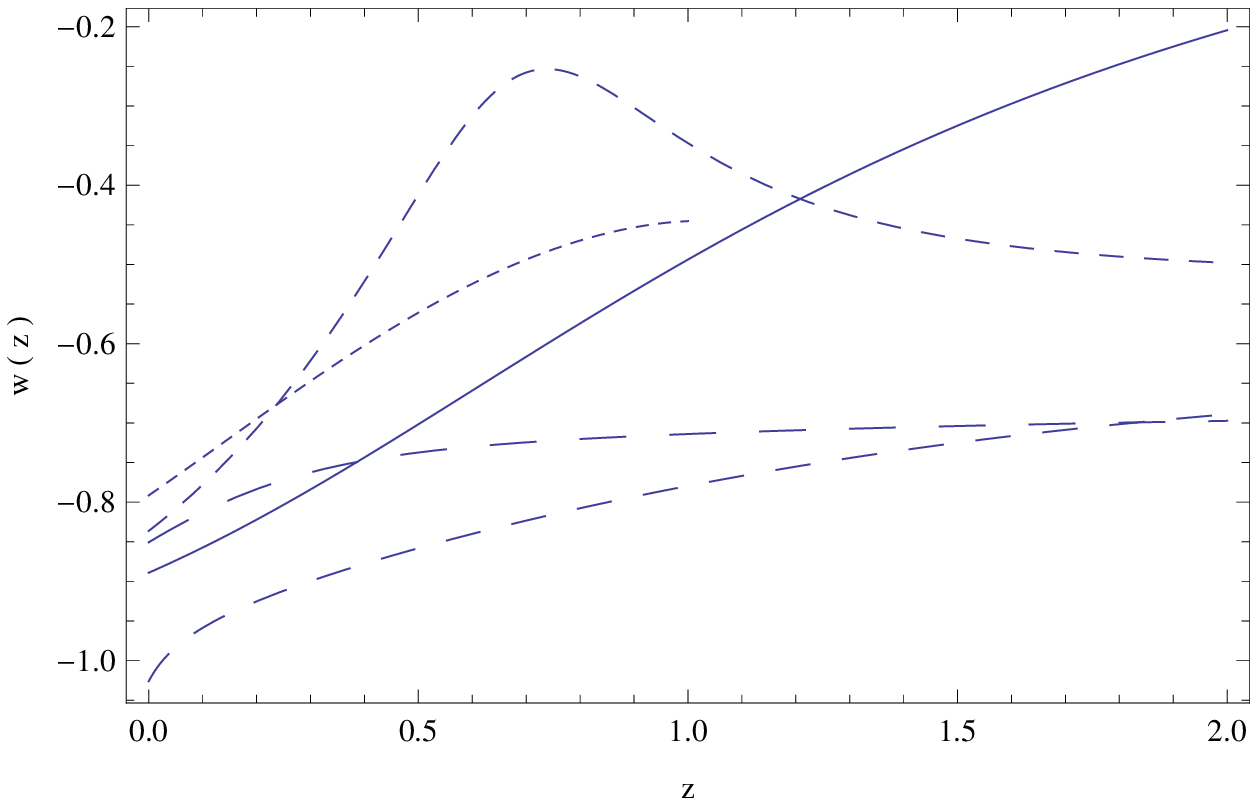}
\caption{Variation of the total equation-of-state parameter
$w$
as a function of $z$ in  the $f(T,\mathcal{T%
})$ gravity theory with $f(T,\mathcal{T})=\protect\alpha \protect\rho
_mT^n+\Lambda $,
for $\alpha _0=-0.0011$, $\lambda =-5.5$, and for five different values of
$n$:
$n=1$ (solid curve), $n=2$ (dotted curve),
 $n=3$ (short-dashed curve),  $n=4$ (dashed
curve), and $n=5$, respectively.}
\label{fig5a}
\end{figure}

Let us now  investigate the effect of $n$ in the function
$f\left(T,\mathcal{T}\right)=\protect\alpha \protect
T^n\,\mathcal{T}+\Lambda=\alpha T^n \rho _m+\Lambda$, on the cosmological
evolution. In order to do so, we fix the values of $\alpha _0 $ and $\lambda $
as $\alpha _0=-0.0011$ and $\lambda =-5.5$, and we consider numerical solutions of
Eqs.~(\ref{eqrA}) and (\ref{eqHAb}) for different values of $n$, by adopting the redshift $z$ as the independent variable. In Figures
\ref{fig1a}-\ref{fig5a} we present the variations with the redshift of the Hubble
function, of the matter energy density, of the
deceleration parameter,  of the dark energy equation-of-state parameter $w_{DE}$, and of the total equation-of-state parameter $w$,
respectively, for $n=1,2,3,4,5$.


Interestingly enough, we observe that even while in the behavior of the Hubble
function, of the scale factor and of the matter energy density there are no major differences between all models with $n\in (1,5)$, the
dynamics of the Universe is very different for different values of $n$, as can be revealed by the
behavior of the deceleration parameter. In particular, while for $n=1$
the Universe starts its evolution from a decelerating phase, followed by an
accelerating one, and ends in an eternally accelerating de Sitter phase, for
$n>1$, all cosmological models begin their
evolution in an accelerating  phase, with $q<0$ at $z=2$,
before entering in a de Sitter exponential expansion ($q=-1$) at $z=0$.
However, the models with $n>1$ exhibit a radical difference in the behavior
of the dark energy sector, which is visible in the evolution of $w_{DE}$.
Specifically, $w_{DE}$ can lie in the quintessence or phantom regime,
depending on the value of $n$. Thus, models that present a similar behavior
in the global dynamics, can be distinguished by the behavior of the
dark energy sector. Nevertheless, note that at late times
$w_{DE}\rightarrow-1$ independently of the value of $n$, and thus in order
to distinguish the various models one should use  $w_{DE}$ at large
redshifts. We mention that, as can be deduced from Eqs.~(\ref{eqrA}) and
(\ref{eqHAb}), independently of   $n$, once the condition $\alpha
(n+1/2)\left(-6 H^2\right)^n \ll 1$ is satisfied, for $\Lambda
\neq 0$ the Universe results in the de Sitter accelerating stage, while for
$\Lambda =0$ its evolution ends in the Einstein--de Sitter, matter-dominated
decelerating phase. Furthermore,  from Fig.~(\ref{fig4a}) and Fig.~(\ref{fig5a}) notice the interesting behavior
that the total equation-of-state parameter $w$ and 
$w_{DE}$ can be either quintessence-like or phantom-like, in all combinations. 
This is easily explained by 
recalling that $w=w_{DE}/\left(1+\rho _m/\rho _{DE}\right)$, and thus according to the signs of $\rho _{DE}$ and
$p_{DE}$ all combinations are possible. Finally, the very similar behaviors  that $w_{DE}$ and $w$ present in some subcases, 
result from the fact that in these subcases $\rho_m\ll\rho_{DE}$, that is the universe is dark-energy dominated.

\subsection{Model B: $f(T,\mathcal{T})=\protect\alpha \mathcal{T}+%
\gamma T^{2}$}

As a second  model describing a simple departure from General Relativity
in the framework of $f(T,\mathcal{T})$ gravity we consider the case
$f(T,\mathcal{T})= \alpha \mathcal{T}+
\gamma T^{2}=\alpha \rho _{m}+\gamma T^{2}=\alpha \rho
_{m}+\beta H^{4}$, where $\alpha $ and $\beta=36\gamma $ are constants. In \
this
case we obtain $f_{T}=\beta T/18=-\beta H^{2}/3$, $f_{TT}=\beta /18$, $f_{%
\mathcal{T}}=\alpha $, and $f_{T\mathcal{T}}=0$, respectively. Thus,
the matter energy density (\ref{eq1}) becomes
\begin{equation}
\rho _{m}=\frac{3\left( 1-\beta H^{2}/2\right) H^{2}}{1+\alpha /2},
\label{1B}
\end{equation}%
while the time variation of the Hubble function (\ref{eq2}) yields
\begin{equation}
\dot{H}=-\frac{3\left( 1+\alpha \right) }{\alpha +2}\frac{\left( 1-\beta
H^{2}/2\right) H^{2}}{1-\beta H^{2}},  \label{2B}
\end{equation}
and therefore, the deceleration parameter (\ref{deccelpar}) is given by
\begin{eqnarray}
q&=&\frac{3\left( 1+\alpha \right) }{\alpha +2}\frac{\left( 1-\beta
H^{2}/2\right) }{1-\beta H^{2}}-1.
\end{eqnarray}
Additionally, the effective dark energy density and pressure, given by Eqs.~(\ref{rhode}) and (\ref{pde}), respectively, can be obtained as
\be\label{528}
\rho _{DE}=\frac{3 H^2 \left(\alpha +\beta  H^2\right)}{\alpha +2},
\ee
\be
p_{DE}=-\frac{3 H^2 \left(\alpha +\beta  H^2\right)}{(\alpha +2) \left(\beta
H^2-1\right)},
\ee
resulting in the following dark energy equation-of-state parameter
\be
w_{DE}=\frac{1}{1-\beta  H^2}.
\ee

In order to examine the behavior of the above observables in a clearer
way, we perform a numerical elaboration of the scenario at hand. We change the independent variable from the time $t$ to the redshift $z$, and we introduce a set of dimensionless variables $\left(h(z),r_m(z), \beta _0\right)$, defined as
\be
H(z)=h(z)H_0, \qquad \rho _m(z)=r_m(z)H_0^2,  \qquad \beta =\frac{\beta _0}{H_0^2}.
\ee

Therefore the basic equations describing the cosmological evolution of the model are
\be
r_m=\frac{3\left( 1-\beta _0h^{2}/2\right) h^{2}}{1+\alpha /2},
\ee
\be\label{zB}
(1+z)h\frac{dh}{dz}=\frac{3\left( 1+\alpha \right) }{\alpha +2}\frac{\left( 1-\beta _0
h^{2}/2\right) h^{2}}{1-\beta _0h^{2}},
\ee
\be
q=\frac{3\left( 1+\alpha \right) }{\alpha +2}\frac{\left( 1-\beta _0
h^{2}/2\right) }{1-\beta _0 h^{2}}-1,
\ee
\be
w_{DE}=\frac{1}{1-\beta _0  h^2}.
\ee
As before, Eq.~(\ref{zB}) must be integrated with the initial condition $h(0)=1$.

\begin{figure}[ht]
\centering
\includegraphics[width=8cm]{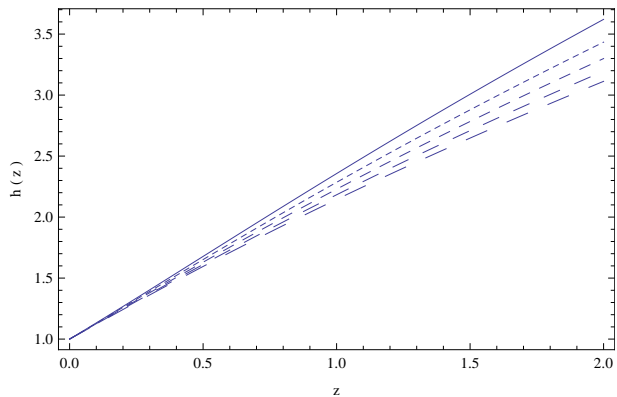}
\caption{Variation of the dimensionless Hubble function $h(z)$ as a function of the redshift $z$
for the model
$f(T,\mathcal{T})=\alpha \mathcal{T}+
\gamma T^{2}=\alpha \rho _m+\beta T^2/36$, with $\beta=36\gamma $, for $\alpha
=-0.15$ and for five
different choices of the parameter $\beta _0 $:
$\beta _0 =-0.10 $ (solid
curve), $\beta _0 =-0.15$ (dotted curve), $%
\beta _0  =-0.20$, (short-dashed curve), $\beta _0 =-0.20$ (dashed
curve), and $\beta _0 =-0.25$ (long-dashed curve), respectively. }
\label{fig6}
\end{figure}

\begin{figure}[!]
\centering
\includegraphics[width=8cm]{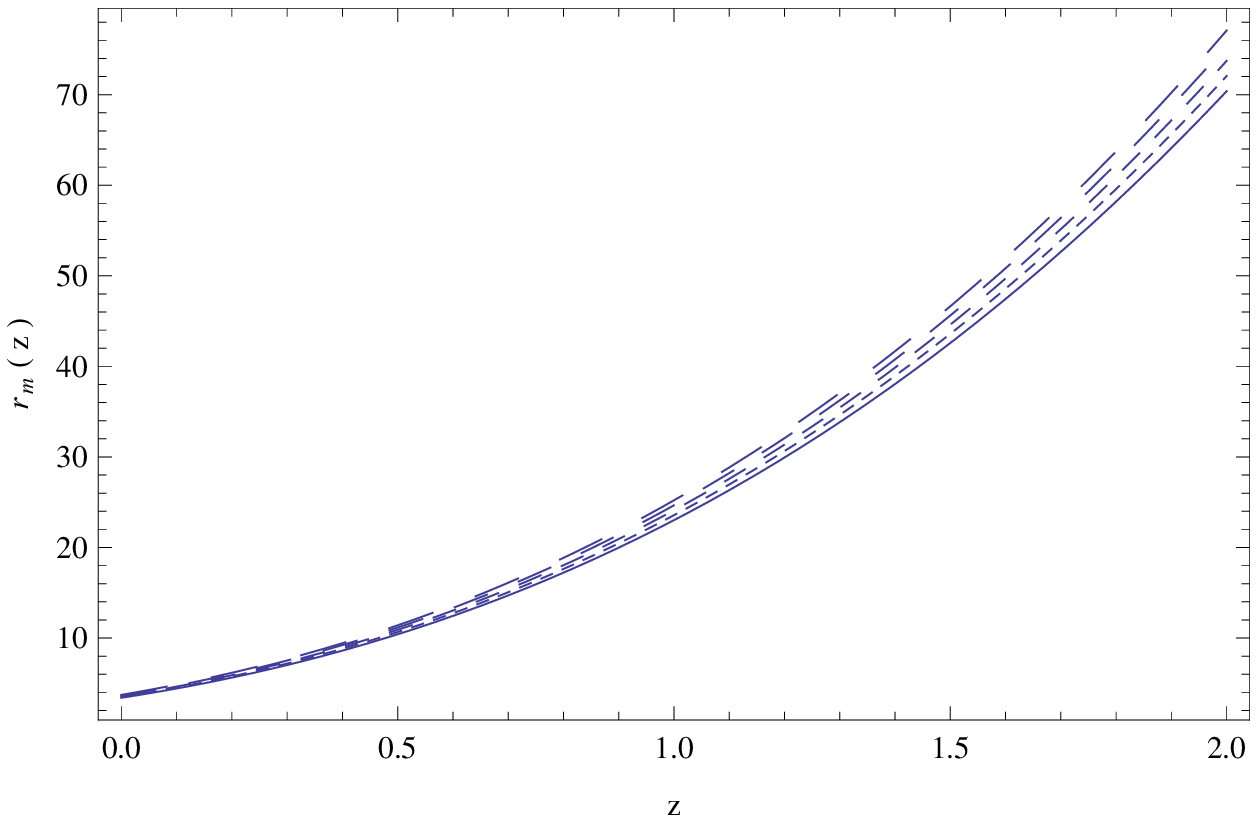}
\caption{Variation of the matter energy density $\protect\rho _m(z)$ as a function of the redshift $z$
for the model
$f(T,\mathcal{T})=\alpha \mathcal{T}+
\gamma T^{2}=\alpha \rho _m+\beta T^2/36$, with $\beta=36\gamma $, for $\alpha
=-0.15$ and  for five
different choices of the parameter $\beta _0 $:
$\beta _0 =-0.10 $ (solid
curve), $\beta _0 =-0.15$ (dotted curve), $%
\beta _0  =-0.20$, (short-dashed curve), $\beta _0 =-0.20$ (dashed
curve), and $\beta _0 =-0.25$ (long-dashed curve), respectively.}
\label{fig8}
\end{figure}

\begin{figure}[!]
\centering
\includegraphics[width=8cm]{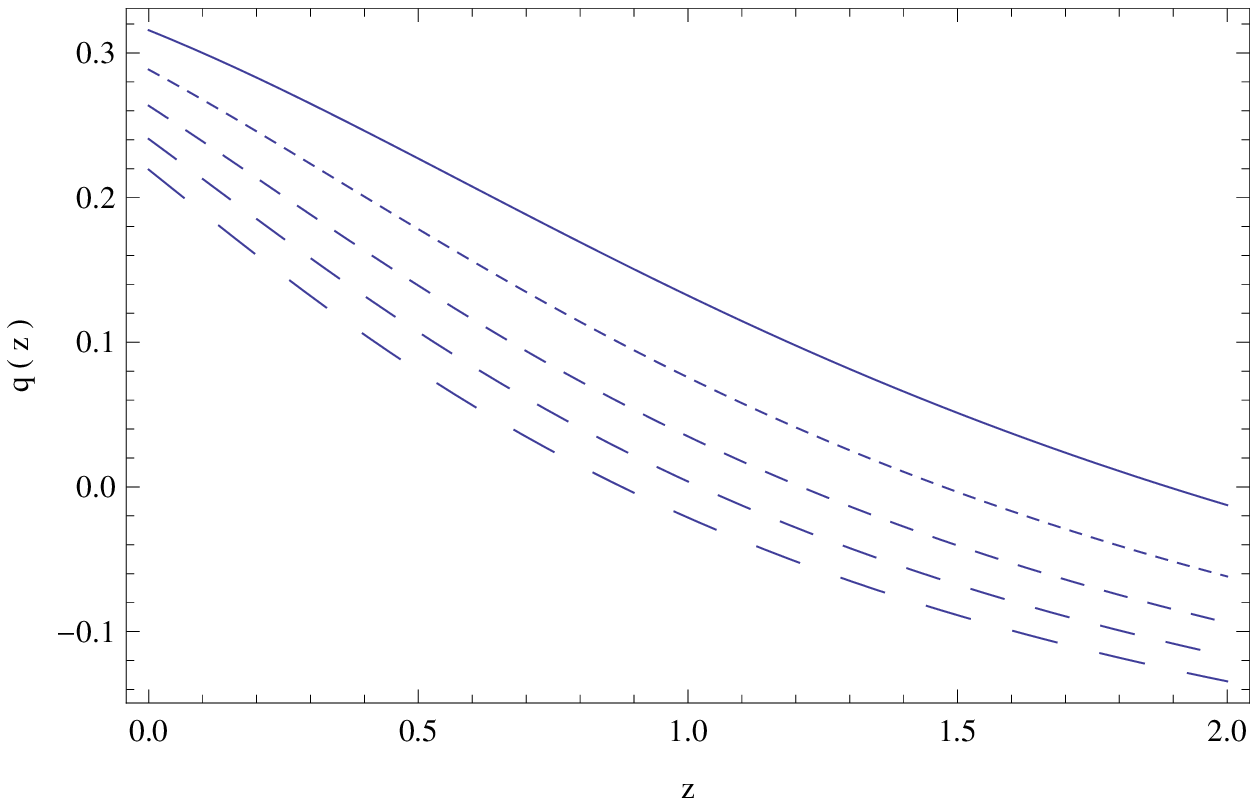}
\caption{Variation of the deceleration parameter $q(z)$ as a function of the redshift $z$ for the model
$f(T,\mathcal{T})=\alpha \mathcal{T}+
\gamma T^{2}=\alpha \rho _m+\beta T^2/36$, with $\beta=36\gamma $, for $\alpha
=-0.15$ and for five
different choices of the parameter $\beta _0 $:
$\beta _0 =-0.10 $ (solid
curve), $\beta _0 =-0.15$ (dotted curve), $%
\beta _0  =-0.20$, (short-dashed curve), $\beta _0 =-0.20$ (dashed
curve), and $\beta _0 =-0.25$ (long-dashed curve), respectively.}
\label{fig9}
\end{figure}

\begin{figure}[!]
\centering
\includegraphics[width=8cm]{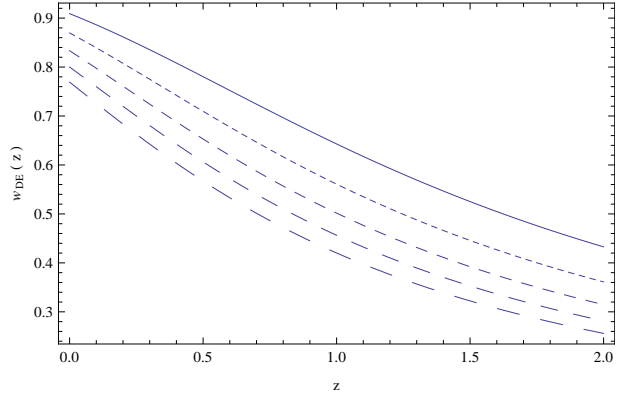}
\caption{Variation of the dark energy equation-of-state parameter
$w_{DE}(z)$ as a function of the redshift $z$ for the model
$f(T,\mathcal{T})=\alpha \mathcal{T}+
\gamma T^{2}=\alpha \rho _m+\beta T^2/36$, with $\beta=36\gamma $, for $\alpha
=-0.15$ and for five
different choices of the parameter $\beta _0 $:
$\beta _0 =-0.10 $ (solid
curve), $\beta _0 =-0.15$ (dotted curve), $%
\beta _0  =-0.20$, (short-dashed curve), $\beta _0 =-0.20$ (dashed
curve), and $\beta _0 =-0.25$ (long-dashed curve), respectively.}
\label{fig10}
\end{figure}

\begin{figure}[!]
\centering
\includegraphics[width=8cm]{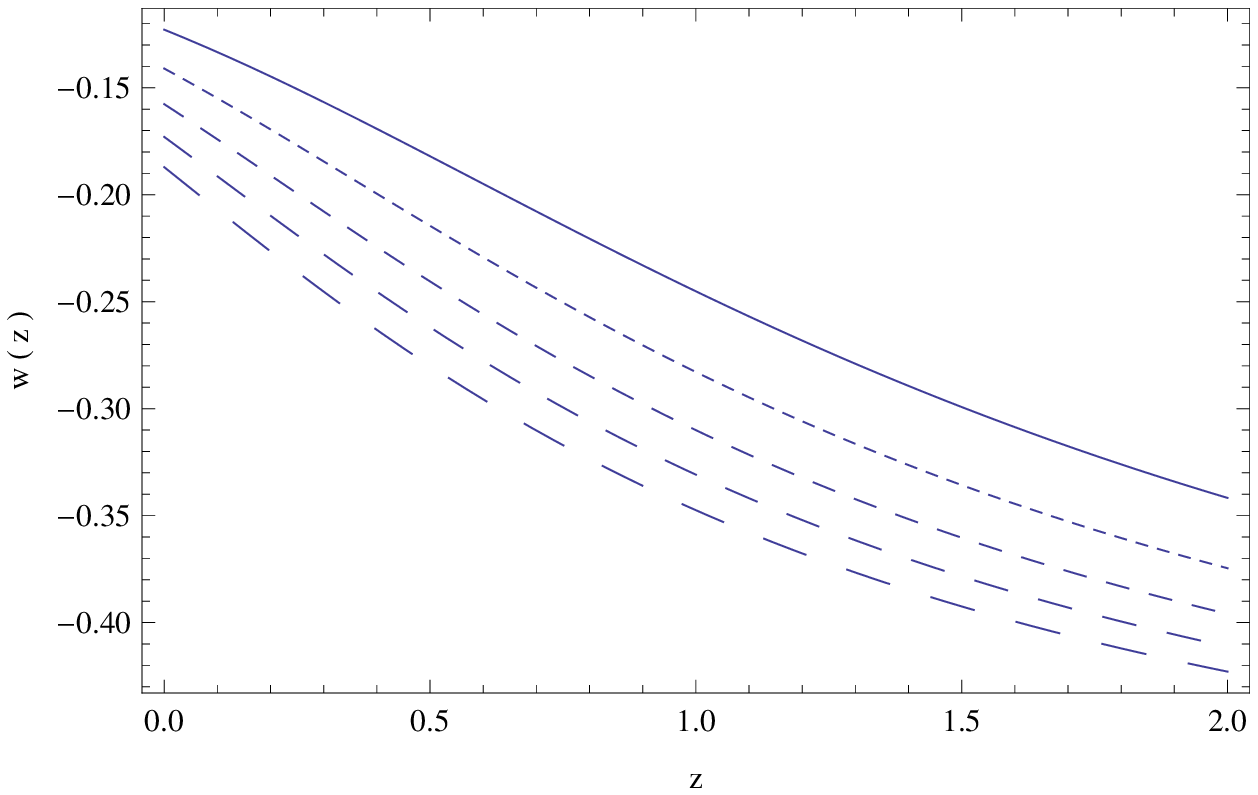}
\caption{Variation of the parameter $w$ of the total  equation-of-state as a function of the redshift $z$ for the model
$f(T,\mathcal{T})=\alpha \mathcal{T}+
\gamma T^{2}=\alpha \rho _m+\beta T^2/36$, with $\beta=36\gamma $, for $\alpha
=-0.15$ and for five
different choices of the parameter $\beta _0 $:
$\beta _0 =-0.10 $ (solid
curve), $\beta _0 =-0.15$ (dotted curve), $%
\beta _0  =-0.20$, (short-dashed curve), $\beta _0 =-0.20$ (dashed
curve), and $\beta _0 =-0.25$ (long-dashed curve), respectively.}
\label{fig10n}
\end{figure}

In Figs.~\ref{fig6}-\ref{fig10n}, we present the evolution of the Hubble function, of the 
matter energy density, of the deceleration parameter, of the dark-energy equation-of-
state parameter, and of the total equation of state parameter, respectively. We mention
that for all these evolutions, we have numerically verified that the
stability conditions extracted in Section \ref{perturb0} are satisfied.

The Hubble function, shown in Fig.~\ref{fig6}, is monotonically decreasing in time (and monotonically increases with the redshift).
 As a result, the matter energy density depicted
in Fig.~\ref{fig8}, decreases monotonically in time. However, the deceleration
parameter $q$, presented in Fig.~\ref{fig9}, exhibits a a large variety of
behaviors, depending on the values of $\alpha $ and $\beta$. In particular,
the Universe can be purely accelerating or purely decelerating, or
experience the transition from deceleration to acceleration.
Finally, the evolution of $w_{DE}$, presented in Fig.~\ref{fig10},
shows that during the entire cosmological evolution
$w_{DE}>0$, tending to $1$ in the large-time limit. A significant difference can be 
observed in the behavior of the parameter of the total equation of state $w$ in Fig. 
\ref{fig10n}, which has an opposite sign as compared to $w_{DE}$, although still in the 
quintessence regime.
This can be explained by our particular choice of the parameters $\alpha <0$ and $\beta<0 $ in Eq.~(\ref{528}), which renders the dark energy density 
 negative during the considered cosmological evolution period. As a result, the ratio $\rho _m/\rho _{DE}<0$, and thus $w<0$. On the other hand $w_{DE}$ is positive for the considered cosmological model, since it is the 
ratio of two negative quantities.

We close this subsection by referring to the limiting behavior of the model
at hand. First of all, the positivity of the matter energy-density implies
that for positive values of $\alpha $ and $\beta $ we must
have $\beta H^{2}/2<1$. Additionally, for $\alpha <-2$ no negative values
of $\beta $ are allowed,
and the Hubble function must satisfy the constraint $\beta H^{2}/2\geq 1$.
For small $H$, that is at late times, and in particular for the time
interval of the cosmological evolution for which $\beta
H^{2}/2 \ll 1$, the Hubble function satisfies the equation $\dot{H}\approx
-3\left( 1+\alpha \right) H^{2}/\left( \alpha +2\right) $, giving $H=\left[
\left( \alpha +2\right) /3\left( 1+\alpha \right) \right] \left( 1/t\right)
$, $a\propto t^{\left( \alpha +2\right) /3\left( 1+\alpha \right) }$, and $%
q\approx \left( 1+2\alpha \right) /\left( \alpha +2\right) $. Thus, the
deceleration parameter is negative for $\alpha \in \left(
-2,-1/2\right) $, however the accelerating phase is not of a de Sitter type,
but it is described by a simple power-law expansion.

\section{Conclusions}\label{Conclusions}

In the present paper, we have introduced a generalization of the $f(T)$
gravitational theory by allowing a general non-minimal coupling between the
torsion scalar $T$ and the trace of the matter energy-momentum tensor
$\mathcal{T}$. The resulting  $f(T,\mathcal{T})$ theory is different
from $f(T)$ gravity, from the curvature-based $f(R,\mathcal{T})$ gravity
\cite{Harko:2011kv}, as well as from the recently constructed nonminimally
torsion-matter coupled theory where $T$ is coupled to the matter Lagrangian
$L_m$ instead of $\mathcal{T}$ \cite{Harko:2014sja}. Therefore, it is a
novel modified gravitational theory. Note that the only restriction imposed on $f$ is the requirement
that it is an analytic function, that is, $f(T,\cal{T})$ is a real function
that is locally given by a convergent power series, and it is infinitely differentiable.

In investigating the physical implications of the theory, in the present
paper we focused on its cosmological implications. The cosmological
equations, obtained for a flat Friedmann-Robertson-Walker type geometry,
are a generalization of both the standard Friedmann equations of General
Relativity, as well as of those of simple $f(T)$ gravity. The coupling
between the torsion scalar and the trace of the matter energy-momentum tensor
contributes with new terms in the effective dark energy density pressure.
More specifically, supplementary terms, proportional to the derivatives of
$f$ with respect to $T$ and $\cal{T}$ appear in the cosmological field
equations. The important feature is that the effective dark energy sector acquires a
contribution from both the $f(T)$ terms, as well as from the matter energy
density and pressure. Due to the extra freedom in the imposed Lagrangian,
$f(T,\mathcal{T})$ cosmology allows for a very wide class of
scenarios and behaviors. Finally, a detailed study of the scalar
perturbations at the linear level reveals that $f(T,\mathcal{T})$ cosmology
can be free of ghosts and instabilities for a wide class of ansatzes and
model parameters.

As applications, we investigated two specific $f(T,\mathcal{T})$ models,
corresponding to simple departures from General Relativity. In
particular, we examined the case where $f$ is chosen to be proportional to the
product of the energy-momentum trace and the torsion scalar at some
power, and the case where $f$ is the sum of the trace of the energy-momentum
tensor and the square of the torsion scalar. We focused on expanding
evolutions, bearing in mind that contracting or bouncing solutions can be also
acquired.

We found a large variety of interesting cosmological behaviors, depending on
the model parameters. For instance, we found specifically evolutions
experiencing a transition from a decelerating to an accelerating state,
capable of describing the late-time cosmic acceleration and the dark energy
epoch. Additionally, we found evolutions where an initial accelerating phase
is followed by a decelerating one, with a subsequent transition to a final
acceleration at late times, a behavior in agreement with the observed thermal
history of the Universe, namely a first inflationary stage, a transition to
non-accelerating, matter-dominated expansion, and then the transition to
late-time accelerating phase. Thus, $f(T,\mathcal{T})$ cosmology offers a
unified description of the universe evolution.

An additional advantage of the scenario at hand, revealing its capabilities,
is that the dark energy equation-of-state parameter can lie in the
quintessence or phantom regime. Moreover, for models with similar
expansion features, $w_{DE}$ may behave very differently, offering a way to
distinguish them. Finally, at late times the universe results either
to a de Sitter exponential expansion, or to eternal power-law accelerated
expansions, with zero matter density, namely, with a complete effective dark-energy
domination.

We close this work by mentioning that the present work is just a first
presentation of $f(T,\mathcal{T})$ gravity and cosmology. In order for this
theory to be a candidate for the description of Nature, many relevant
investigations are necessary. In particular one should perform a detailed
comparison with  cosmological observations (for instance using data from
 Type Ia Supernovae (SNIa), Baryon Acoustic Oscillations (BAO), and Cosmic
Microwave Background (CMB), along with requirements of Big Bang
Nucleosynthesis (BBN)), which could constrain the allowed
ansatzes and parameter ranges. Furthermore, after extracting the spherically
symmetric solutions, one could confront $f(T,\mathcal{T})$  gravity with
Solar System data. Additionally, one could use the scalar perturbation
equations extracted in the present work in order to perform a detailed
confrontation with the growth-index data. Moreover, one could extend the
perturbation analysis to the vector and tensor modes, and use them in order
to predict the inflationary induced tensor-to-scalar ratio, especially under
the recent BICEP2 measurements that can exclude a large class of models
\cite{BICEP2}. These necessary studies lie beyond the scope of the present
work, and are left for a separate project.

\begin{acknowledgments}
We would like to thank the anonymous referee for comments and suggestions that helped us to significantly improve our manuscript. We are grateful to Prof. Kourosh Nozari for calling to our attention Ref. \cite{Kiani:2013pba}, after our paper was submitted.
FSNL acknowledges financial  support of the Funda\c{c}\~{a}o para a
Ci\^{e}ncia e Tecnologia through an Investigador FCT Research contract, with
reference IF/00859/2012, funded by FCT/MCTES (Portugal), and grants
CERN/FP/123618/2011 and EXPL/FIS-AST/1608/2013. GO would like to thank CAPES and
FAPEMIG for financial support. The research of ENS is implemented within the
framework of the Operational Program ``Education and Lifelong Learning''
(Actions Beneficiary: General Secretariat for Research and Technology), and
is co-financed by the European Social Fund (ESF) and the Greek State.
\end{acknowledgments}

\appendix

\section{Coefficients of the stability equation}
\label{Coefficients}

In this appendix, we give the coefficients $\Gamma$, $\mu^2$, $c_s^2$ and $D$
of the perturbation equation (\ref{phiddk}):
\begin{eqnarray}
\label{phiddk22}
\ddot{\tilde{\phi}}_k+\Gamma \dot{\tilde{\phi}}_k+\mu^2
\tilde{\phi}_k+c_s^2\frac{k^2}{a^2}
\tilde{\phi}_k=D.
\end{eqnarray}
Concerning the effective mass we have
\begin{equation}
\label{omega2}
 \mu^2=\mu^2_{(1)}+\mu^2_{(2)}+\mu^2_{(3)}+\mu^2_{(4)}+\mu^2_{(5)}+
\mu^2_{(6)},
\end{equation}
with
\begin{eqnarray}
 \mu^2_{(1)}&=&2\dot{H}\left(f_{T}+1\right)   \frac{
B}{E}
+2H   \left(\dot{\rho}_{m}-3 \dot{p}_{m}
\right) f_{T\mathcal{T}}\frac{B^2}{F}
\nonumber\\
&&
 +H^2     \Big[4 \dot{H} \left(3 A f_{T\mathcal{T}
}-5 B f_{TT}\right) \frac{3 B}{F}
+A f_{\mathcal{T}}+B
f_{T}+B\Big] ,
\end{eqnarray}
\begin{eqnarray}
\mu^2_{(2)}&=&\frac{12 H^3}{F} \Big\{8
\dot{p}_{m}f_{T\mathcal{T}}\left(B f_{T\mathcal{T}}-A
f_{\mathcal{T}\mathcal{T}}\right)
+ \left(\dot{\rho}_{m}- 3\dot{p}_{m}\right)
 \left[2 p_{m} f_{T\mathcal{T}}
\left(B f_{
T\mathcal{T}\mathcal{T}}-A f_{\mathcal{T}\mathcal{T}\mathcal{T}}
\right)\right.\notag \\
&&
\!\!\!\!\!\!\!\!\!\! \!\!\!\!\!\!\!\!\!\!  \!\!\!\!\!\!
+A B f_{T\mathcal{T}\mathcal{T}}-2 A \rho_{m} f_{\mathcal{T}\mathcal{T}
\mathcal{T}} f_{T\mathcal{T}}-3 A f_{\mathcal{T}\mathcal{T}} f_{T\mathcal
{T}}
\left.
-B^2 f_{TT\mathcal{T}}+2 B \rho_{m} f_{T\mathcal{T}}
f_{T\mathcal{T}
\mathcal{T}}+3 B f_{T\mathcal{T}}^2\right]
\Big\},
\end{eqnarray}
\begin{eqnarray}
\mu^2_{(3)}&=&\frac{12 H^4}{F} \Big\{4
f_{T\mathcal{T}}^2 \left[9 A
\dot{H}+B (p_{m}+\rho_{m})\right]
+f_{T\mathcal{T}} \Big\{
 2 B( f_{T}+1)+A B
\notag \\
&&
 +24 \dot{H}
\Big[(p_{m}+\rho_{m}) (A f_{T\mathcal{T}\mathcal{T}}-B f_{TT\mathcal{T}}
)-3 B f_{TT}\Big]
\Big\}\notag \\
&&
 -3 B \Big[4 \dot{H} (A f_{TT\mathcal{T}}-B f_
{TTT})+f_{TT} (f_{\mathcal{T}}+B)\Big] \Big\},
\end{eqnarray}
\begin{eqnarray}
\mu^2_{(4)}&=&\frac{144 H^5}{F}
\Big\{8 \dot{p}_{m} f_{TT}   f_{T\mathcal{T}\mathcal{T}}
f_{T\mathcal{T}}
+
\left(\dot{\rho}_{m}-3 \dot{p}_{m}  \right)
\big\{f_{T\mathcal{T}}
 \left\{A f_{T\mathcal{T}\mathcal{T}}-B
f_{TT\mathcal{T}}\right. \nonumber\\
 && \left.
 +f_{TT}
\left[2
(p_{m} +\rho_{m}) f_{\mathcal{T}\mathcal{T}\mathcal{T}}+3
f_{\mathcal{T}\mathcal{T }}\right]
\right\}
-B f_{T\mathcal{T}\mathcal{T}} f_{TT}\big\}
\Big\},
\end{eqnarray}
\begin{eqnarray}
\mu^2_{(5)}&=& -\frac{432 H^6}{F} \Big\{f_{T\mathcal{T}}
\left\{4 \dot{H}
(A f_{TT\mathcal{T}}-B f_{TTT})
+f_{TT} \left[8 \dot{H}
(p_{m}+\rho_{m}) f_{ T\mathcal{T}\mathcal{T}}+B\right]
\right\}\nonumber\\
&&
-4 B \dot{H} f_{TT} f_{TT\mathcal{T}}
+12 \dot{H} f_{TT} f_{T\mathcal{T}}^2\Big\},
\end{eqnarray}
and
\begin{eqnarray}
\mu^2_{(6)}&=&\frac{1728}{F}  H^7 f_{T\mathcal{T}} f_{TT}
\left[12 H \dot{H} f_{TT\mathcal{T}}
+\left(3
\dot{p}_{m}-\dot{\rho}_{m} \right) f_{T\mathcal{T}\mathcal{T}}\right],
 \end{eqnarray}
respectively.

Concerning the sound speed, we have
\begin{eqnarray}
\label{omega2b}
c_s^2=c^2_{s(1)}+c^2_{s(2)}+c^2_{s(3)}+c^2_{s(4)},
\end{eqnarray}
with the following relations
\begin{equation}
c^2_{s(1)}=\frac{\left(f_{T}+1\right)}{E}
\left(4 \dot{H} f_{T\mathcal{T}}+f_{\mathcal{T}}\right),
\end{equation}
\begin{eqnarray}
c^2_{s(2)}&=&\frac{4 H}{F} \Big\{   B \left(\dot{\rho}_{m}-3 \dot{p}_{m}\right)
(f_{T}+1) f_{
T\mathcal{T}\mathcal{T}}
+f_{T\mathcal{T}}
\left\{
B
\left(\dot{\rho}_{m}-3 \dot{p}_{m}\right)
f_{T\mathcal{T}}
\right.  \notag \\
&&\left.
+\left(f_{T} +1\right)
\left[\left(\dot{p}_{m}-3 \dot{\rho}_{m}\right) f_{\mathcal{T}
\mathcal{T}}\right.
\left.
+ 2 (p_{m}+\rho_{m}) (3 \dot{p}_{m}-\dot{\rho}_{m}) f_{\mathcal{T}\mathcal{
T}\mathcal{T}}\right]
\right\}
\Big\},
\end{eqnarray}
\begin{eqnarray}
c^2_{s(3)}&=&\frac{4H^2}{F}
\Big\{
-12 B \dot{H}\left[ f_{TT}f_{T\mathcal{T}}
+ (f_{T}+1) f_{TT\mathcal{T}}\right]
\notag
\\
&&
 +
  f_{T\mathcal{T}}(f_{T}+1) \big\{12 \dot{H} \left[2 (p_{m}+\rho_{m})
f_{T\mathcal{T}
\mathcal{T}}
+3 f_{T\mathcal{T}}\right]
+B\big\} \Big\},
\end{eqnarray}
\begin{eqnarray}
\!\!\!  \!\!\!  \!\!\!
c^2_{s(4)}=-\frac{48 H^3}{F}     \left(3
\dot{p}_{m}-\dot{\rho}_{m}\right)
(f_{T}+1) f_{T\mathcal{T}} f_{T\mathcal{T}\mathcal{T}},
\end{eqnarray}
respectively.

Concerning the frictional coefficient we have
\begin{eqnarray}
\Gamma=\Gamma^{(1)}+\Gamma^{(2)}+\Gamma^{(3)}+\Gamma^{(4)}+\Gamma^{(5)}
+\Gamma^{(6)},
\end{eqnarray}
with
\begin{eqnarray}
\Gamma^{(1)}=
\frac{f_{T\mathcal{T}}}{F} \Big[20736\, H^7 \dot{H} f_{TT}
f_{TT\mathcal{T}}
-\left(3 \dot{p}_{m}-\dot{\rho}_{m}\right) \left(B^2-1728
H^6 f_{T\mathcal{T}\mathcal{T}} f_{TT}\right)\Big],
\end{eqnarray}
\begin{eqnarray}
\Gamma^{(2)}=\frac{H}{E} \left\{4 \left[\dot{H} \left(6 A f_{T\mathcal{T}
}-9 B f_{TT}\right)+B f_{T}+B\right]
+3 A f_{\mathcal{T}}+\frac{4
k^2}{
a^2} \left(f_{T}+1\right) f_{T\mathcal{T}}\right\},
\end{eqnarray}
\begin{eqnarray}
&&
\Gamma^{(3)}= \frac{\mu^2_{(2)}}{H},
 \end{eqnarray}
\begin{eqnarray}
 \Gamma^{(4)}&=&\frac{12 H^3}{F}
 \Big\{36 A \dot{H} f_{T\mathcal{T}}^2+4 A B  f_{T\mathcal{T}}
 - B \big[12 \dot{H} \left(A f_{TT\mathcal{T}}-B
f_{TTT}\right)
\nonumber\\
&&
\!\!\! \!
+f_{TT}  \left(3 f_{\mathcal{T}}+4 B\right)\big]
- 12 \dot{H} f_{T\mathcal{T}}   \left[5 B
f_{TT}
+2 (p_{m}+\rho_{m}) \left(B f_{
TT\mathcal{T}}   -A
f_{T\mathcal{T}\mathcal{T}}\right)\right]
\Big\},
 \end{eqnarray}
 \begin{eqnarray}
\Gamma^{(5)}=\frac{\mu^2_{(4)}}{H},
  \end{eqnarray}
  and
  \begin{eqnarray}
\Gamma^{(6)}&=&\frac{576 H^5}{F} \Big\{
3 B \dot{H} f_{TT} f_{TT\mathcal{T}}-9 \dot{H} f_{TT} f_{T\mathcal{T}}
^2
-f_{T\mathcal{T}} \left[3 \dot{H}
(A f_{TT\mathcal{T}}-B f_{TTT})\right.
\nonumber\\
&&
\left.+ 6 \dot{H} (p_{m}+\rho_{m})
f_{TT} f_{ T\mathcal{T}\mathcal{T}}+Bf_{TT}  \right]
\Big\},
  \end{eqnarray}
  respectively.

The coefficient $D$ of the right-hand side of (\ref{phiddk22}) is given by
\begin{eqnarray}
D=-D_{1} \delta{\dot{\tilde{p}}^{k}_{m}}-D_{2} \delta{\tilde{p}}^{k}_{m},
\end{eqnarray}
where
\begin{eqnarray}
 D_{1}=\frac{H  f_{T\mathcal{T}}}{E} (I+36 H^2 f_{T\mathcal{T}}),
\end{eqnarray}
and
\begin{eqnarray}
 D_{2}=D_{2}^{(1)}+D_{2}^{(2)}+D_{2}^{(3)}+D_{2}^{(4)}+D_{2}^{(5)}+D_{2}^{(6)
},
\end{eqnarray}
with
\begin{eqnarray}
D_{2}^{(1)}&=&\frac{1}{4 E} \left[(4 \dot{H} f_{T\mathcal{T}}+f_{\mathcal{T
}}) I-16 \pi G B\right],
\end{eqnarray}
\begin{eqnarray}
D_{2}^{(2)}&=&-\frac{H}{F}
\Big\{
\dot{\rho}_{m} \left\{f_{T\mathcal{T}} \left[f_{\mathcal{T}\mathcal{T}} (3
I-8 B)\right.
\left.
 +2 I (p_{m}+\rho_{m})
f_{\mathcal{T}\mathcal{T}\mathcal{T}}\right]
-BI
f_{T\mathcal{T}\mathcal{T}}\right\}
\notag
\\
&&  -\dot{p}_{m} \left\{f_{T\mathcal{T}} \left[f_{
\mathcal{T}\mathcal{T}} (I-24 B)\right.
\left.
+6 I (p_{m}+\rho_{m}) f_{\mathcal{T}
\mathcal{T}\mathcal{T}}\right]
-3 B I f_{T\mathcal{T}\mathcal{T}}\right\}
\Big\},
\end{eqnarray}
\begin{eqnarray}
D_{2}^{(3)}&=&\frac{3
H^2}{F}
\Big\{
4 \dot{H}\left[ f_{T\mathcal{T}}^2 (3 I-5
B)- I B f_{
TT\mathcal{T}}\right]
+f_{T\mathcal{T}}
\big\{8 f_{\mathcal{T}} \big[5 \dot{H}
(p_{m}+\rho_{m})
f_{T\mathcal{T}\mathcal{T}}+B\big] \notag \\
&&
+3 \left[B-2 (p_{m}+\rho_{m})
f_{\mathcal{T}\mathcal{T}}\right]
\big[8 \dot{H}
(p_{m}
+\rho_{m}) f_{T\mathcal{T}\mathcal{T}}+B\big] \big\}
\Big\},
\end{eqnarray}
\begin{eqnarray}
    D_{2}^{(4)}&=&\frac{12 H^3
f_{T\mathcal{T}}}{F}
\Big\{(\dot{\rho}_{m}-3
\dot{p}_{m}) f_{T\mathcal{T}\mathcal{T}} (I+3 B)
+3 f_{T\mathcal{T}} \big[(\dot{p}_{m}-3 \dot{\rho}_{m})
f_{\mathcal{T}\mathcal{T}}
\nonumber\\
&& +
2 (p_{m}+\rho_{m}) (3 \dot{p}_{m}-\dot{\rho}_{m}) f_{\mathcal{T}\mathcal{
T}\mathcal{T}}\big]\Big\},
\end{eqnarray}
\begin{equation}
 D_{2}^{(5)}=\frac{36 H^4 f_{T\mathcal{T}}}{F}
\Big\{ 3 f_{T\mathcal{T}}  B -4 \dot{H}
f_{TT\mathcal{T}} \left(I+3
B\right)
+3 f_{T\mathcal{T}}
\big\{4 \dot{H} \left[2 (p_{m}+\rho_{m})
f_{T\mathcal{T}\mathcal{T}}+3 f_
{T\mathcal{T}}\right]
\big\}
\Big\},
\end{equation}
and
\begin{eqnarray}
    D_{2}^{(6)}&=&-\frac{432 H^5
f_{T\mathcal{T}}^2}{F} \Big[12 H \dot{H}
f_{TT\mathcal{T}}
+\left(3 \dot{p}_{m}-\dot{\rho}_{m}\right)
f_{T\mathcal{T}
\mathcal{T}}\Big],
 \end{eqnarray}
 respectively.

 Finally, in all the above expressions we have introduced the coefficients
\begin{eqnarray}
 &&A\equiv 2 \left(p_{m}+\rho_{m}\right) f_{T\mathcal{T}}+f_{T}+1,\nonumber\\
&&
 B\equiv
2 \left[8 \pi G+\left(p_{m}+\rho_{m}\right) f_{\mathcal{T
}\mathcal{T}}-6 H^2 f_{T\mathcal{T}}\right]+f_{\mathcal{T}},
\nonumber\\
&& E\equiv
12 H^2 \left[A f_{T\mathcal{T}}-f_{TT} \left(12 H^2 f_{T\mathcal{T}}
+B\right)\right] +B \left(f_{T}+1\right),\nonumber\\
&&
 I\equiv-6 (p_{m}+\rho_{m}) f_{\mathcal{T}\mathcal{T}}+5 f_{\mathcal{T}}
+3 B,\nonumber\\
&&
F\equiv B E,
\end{eqnarray}
respectively.


\begin{thebibliography}{99}

\bibitem{acc}
  M.~Betoule {\it et al.}  [SDSS Collaboration],
  {\it{Improved cosmological constraints from a joint analysis of the SDSS-II and SNLS supernova samples}},
      [\href{http://xxx.lanl.gov/abs/1401.4064}
{{\tt arXiv:1401.4064}}].

 
\bibitem{Copeland:2006wr}
  E.~J.~Copeland, M.~Sami and S.~Tsujikawa,
  {\it{Dynamics of dark energy}},
  Int.\ J.\ Mod.\ Phys.\  D {\bf 15}, 1753 (2006),
     [\href{http://xxx.lanl.gov/abs/hep-th/0603057}
{{\tt arXiv:hep-th/0603057}}].


\bibitem{Cai:2009zp}
  Y.~-F.~Cai, E.~N.~Saridakis, M.~R.~Setare and J.~-Q.~Xia,
  {\it{Quintom Cosmology: Theoretical implications and observations}},
  Phys.\ Rept.\  {\bf 493}, 1 (2010),
       [\href{http://xxx.lanl.gov/abs/0909.2776}
{{\tt arXiv:0909.2776}}].



\bibitem{Capozziello:2011et}
  S.~Capozziello and M.~De Laurentis,
  {\it{Extended Theories of Gravity}},
  Phys.\ Rept.\  {\bf 509}, 167 (2011),
  [\href{http://xxx.lanl.gov/abs/1108.6266}
{{\tt arXiv:1108.6266}}].





\bibitem{DeFelice:2010aj}
  A.~De Felice and S.~Tsujikawa,
       {\it{f(R) theories}},
  Living Rev.\ Rel.\  {\bf 13}, 3 (2010),
    [\href{http://xxx.lanl.gov/abs/1002.4928}
{{\tt arXiv:1002.4928}}].

\bibitem{Nojiri:2010wj}
  S.~'i.~Nojiri and S.~D.~Odintsov,
      {\it{Unified cosmic history in modified gravity: from F(R) theory to
Lorentz non-invariant models}},
  Phys.\ Rept.\  {\bf 505}, 59 (2011),
      [\href{http://xxx.lanl.gov/abs/1011.0544}
{{\tt arXiv:1011.0544}}].

  
\bibitem{Lobo:2008sg} 
  F.~S.~N.~Lobo,
  {\it{The Dark side of gravity: Modified theories of gravity}},
  Dark Energy-Current Advances and Ideas, {\bf 173-204},   Research Signpost, ISBN 978 (2009),
        [\href{http://xxx.lanl.gov/abs/0807.1640}
{{\tt arXiv:0807.1640}}].
 
  
  
\bibitem{Sahni:2006pa}
  V.~Sahni and A.~Starobinsky,
  {\it{Reconstructing Dark Energy}},
  Int.\ J.\ Mod.\ Phys.\ D {\bf 15}, 2105 (2006),
       [\href{http://xxx.lanl.gov/abs/astro-ph/0610026}
{{\tt arXiv:astro-ph/0610026}}].



\bibitem{Unzicker:2005in}
 A.~Unzicker and T.~Case,
{\it{Translation of Einstein's attempt of a unified field
theory with teleparallelism}},
[\href{http://xxx.lanl.gov/abs/physics/0503046}
{{\tt arXiv:physics/0503046}}].




\bibitem{TEGR} 
   C. M\"{o}ller,
   {\it{Conservation laws and absolute parallelism in general relativity}},
    Mat. Fys. Skr. Dan. Vid. Selsk. {\bf 1}, 3
(1961).

\bibitem{TEGR22} 
  C. Pellegrini and J. Plebanski,
   {\it{Tetrad fields and gravitational fields}},
   Mat. Fys. Skr. Dan. Vid. Selsk. {\bf 2}, 1
(1963).


\bibitem{Hayashi:1979qx}
  K.~Hayashi and T.~Shirafuji,
     {\it{New general relativity}},
  Phys.\ Rev.\  D {\bf 19}, 3524 (1979)
  [Addendum-ibid.\  D {\bf 24}, 3312 (1982)].




    \bibitem{JGPereira}
R. Aldrovandi and J. G. Pereira, {\it Teleparallel Gravity: An Introduction},
Springer, Dordrecht (2013).





 \bibitem{Arcos:2005ec}
  H.~I.~Arcos and J.~G.~Pereira,
   {\it{Torsion Gravity: a Reappraisal,}}
  Int.\ J.\ Mod.\ Phys.\ D \textbf{13}, 2193 (2004),
  [\href{http://xxx.lanl.gov/abs/gr-qc/0501017}
{{\tt arXiv:gr-qc/0501017}}].


 
  
\bibitem{Maluf:2013gaa}
  J.~W.~Maluf,
  {\it{The teleparallel equivalent of general relativity}},
  Annalen Phys.\  {\bf 525}, 339 (2013),
     [\href{http://xxx.lanl.gov/abs/1303.3897}
{{\tt arXiv:1303.3897}}].



\bibitem{Pereira:2013qza} 
  J.~G.~Pereira,
 {\it {Teleparallelism: A New Insight Into Gravity}},
  in Springer Handbook of Spacetime, edited by A. Ashtekar and V. Petkov (Springer, Dordrecht, 2013),
       [\href{http://xxx.lanl.gov/abs/1302.6983}
{{\tt arXiv:1302.6983}}].

 
  
  
  

\bibitem{Ferraro:2006jd}
  R.~Ferraro and F.~Fiorini,
   {\it{Modified teleparallel gravity: Inflation without inflaton}},
  Phys.\ Rev.\ D {\bf 75}, 084031 (2007),
  [\href{http://xxx.lanl.gov/abs/gr-qc/0610067}
{{\tt arXiv:gr-qc/0610067}}].


  
\bibitem{Ben09}
  G.~R.~Bengochea and R.~Ferraro,
  {\it{Dark torsion as the cosmic speed-up}},
  Phys.\ Rev.\ D \textbf{79}, 124019 (2009),
 [\href{http://xxx.lanl.gov/abs/0812.1205}
 {{\tt arXiv:0812.1205}}].

 
 
 

\bibitem{Linder:2010py}
  E.~V.~Linder,
     {\it{Einstein's Other Gravity and the Acceleration of the Universe}},
  Phys.\ Rev.\  D {\bf 81}, 127301 (2010),
[\href{http://xxx.lanl.gov/abs/1005.3039}
{{\tt arXiv:1005.3039}}].




\bibitem{Chen:2010vaoooo}
S.~H.~Chen, J.~B.~Dent, S.~Dutta and E.~N.~Saridakis,
\textit{Cosmological perturbations in f(T) gravity},
Phys.\ Rev.\ D
\textbf{ 83}, 023508 (2011),
[\href{http://xxx.lanl.gov/abs/1008.1250}
{\texttt{%
arXiv:1008.1250}}]. 


\bibitem{Dent:2011zz}
J.~B.~Dent, S.~Dutta, E.~N.~Saridakis,
  {\it{f(T) gravity mimicking dynamical dark energy. Background and
perturbation analysis}},
  JCAP {\bf 1101}, 009 (2011)
[\href{http://xxx.lanl.gov/abs/1010.2215}
{{\tt arXiv:1010.2215}}].


 



\bibitem{Chen:2010va}
  P.~Wu, H.~W.~Yu,
  {\it{Observational constraints on $f(T)$ theory}},
  Phys.\ Lett.\ \textbf{B693}, 415 (2010),
 [\href{http://xxx.lanl.gov/abs/1006.0674}
 {{\tt arXiv:1006.0674}}].
 



\bibitem{Zheng:2010am}
  R.~Zheng and Q.~G.~Huang,
    {\it{Growth factor in f(T) gravity}},
  JCAP {\bf 1103}, 002 (2011),
 [\href{http://xxx.lanl.gov/abs/1010.3512}
 {{\tt arXiv:1010.3512}}].

\bibitem{Bamba:2010wb}
  K.~Bamba, C.~Q.~Geng, C.~C.~Lee and L.~W.~Luo,
      {\it{Equation of state for dark energy in $f(T)$ gravity}},
  JCAP {\bf 1101}, 021 (2011),
 [\href{http://xxx.lanl.gov/abs/1011.0508}
 {{\tt arXiv:1011.0508}}].

\bibitem{Cai:2011tc}
  Y.~-F.~Cai, S.~-H.~Chen, J.~B.~Dent, S.~Dutta, E.~N.~Saridakis,
  {\it{Matter Bounce Cosmology with the f(T) Gravity}},
  Class.\ Quant.\ Grav.\  {\bf 28}, 2150011 (2011),
 [\href{http://xxx.lanl.gov/abs/1104.4349}
 {{\tt arXiv:1104.4349}}].


\bibitem{Sharif001}
  M.~Sharif, S.~Rani,
  {\it{F(T) Models within Bianchi Type I Universe}},
  Mod.\ Phys.\ Lett.\  {\bf A26}, 1657 (2011),
 [\href{http://xxx.lanl.gov/abs/1105.6228}
 {{\tt arXiv:1105.6228}}].

\bibitem{Li:2011rn}
  M.~Li, R.~X.~Miao and Y.~G.~Miao,
     {\it{Degrees of freedom of $f(T)$ gravity}},
  JHEP {\bf 1107}, 108 (2011),
 [\href{http://xxx.lanl.gov/abs/1105.5934}
 {{\tt arXiv:1105.5934}}].

\bibitem{Capozziello:2011hj}
  S.~Capozziello, V.~F.~Cardone, H.~Farajollahi and A.~Ravanpak,
     {\it{Cosmography in f(T)-gravity}},
  Phys.\ Rev.\  D {\bf 84}, 043527 (2011),
 [\href{http://xxx.lanl.gov/abs/1108.2789}
 {{\tt arXiv:1108.2789}}].

 
 
\bibitem{Daouda:2011rt}
  M.~H.~Daouda, M.~E.~Rodrigues and M.~J.~S.~Houndjo,
     {\it{Static Anisotropic Solutions in f(T) Theory}},
[\href{http://xxx.lanl.gov/abs/1109.0528}
{{\tt arXiv:1109.0528}}].



  
\bibitem{Wu:2011kh}
  Y.~P.~Wu and C.~Q.~Geng,
     {\it{Primordial Fluctuations within Teleparallelism}},
     Phys.\ Rev.\ D {\bf 86}, 104058 (2012),
[\href{http://xxx.lanl.gov/abs/1110.3099}
{{\tt arXiv:1110.3099}}].

\bibitem{Wei:2011aa}
  H.~Wei, X.~J.~Guo and L.~F.~Wang,
     {\it{Noether Symmetry in $f(T)$ Theory}},
  Phys.\ Lett.\  B {\bf 707}, 298 (2012),
 [\href{http://xxx.lanl.gov/abs/1112.2270}
 {{\tt arXiv:1112.2270}}].

\bibitem{Atazadeh:2011aa}
  K.~Atazadeh and F.~Darabi,
     {\it{$f(T)$ cosmology via Noether symmetry}},
 Eur.Phys.J. C72 (2012) 2016,
 [\href{http://xxx.lanl.gov/abs/1112.2824}
 {{\tt arXiv:1112.2824}}].

 
\bibitem{Farajollahi:2011af}
  H.~Farajollahi, A.~Ravanpak and P.~Wu,
     {\it{Cosmic acceleration and phantom crossing in $f(T)$-gravity}},
  Astrophys.\ Space Sci.\  {\bf 338}, 23 (2012),
 [\href{http://xxx.lanl.gov/abs/1112.4700}
 {{\tt arXiv:1112.4700}}].

 
 
\bibitem{Karami:2012fu}
  K.~Karami and A.~Abdolmaleki,
   {\it{Generalized second law of thermodynamics in f(T)-gravity}},
 JCAP 1204 (2012) 007,
 [\href{http://xxx.lanl.gov/abs/1201.2511}
 {{\tt arXiv:1201.2511}}].

\bibitem{Iorio:2012cm}
  L.~Iorio and E.~N.~Saridakis,
      {\it{Solar system constraints on f(T) gravity}},
 Mon.Not.Roy.Astron.Soc. 427 (2012) 1555,
 [\href{http://xxx.lanl.gov/abs/1203.5781}
 {{\tt arXiv:1203.5781}}].

\bibitem{Cardone:2012xq}
  V.~F.~Cardone, N.~Radicella and S.~Camera,
    {\it{Accelerating f(T) gravity models constrained by recent cosmological
 data}},
  Phys.\ Rev.\ D {\bf 85}, 124007 (2012),
   [\href{http://xxx.lanl.gov/abs/1204.5294}
 {{\tt arXiv:1204.5294}}].

 \bibitem{Jamil:2012ti}
  M.~Jamil, D.~Momeni and R.~Myrzakulov,
    {\it{Wormholes in a viable f(T) gravity}},
  Eur.\ Phys.\ J.\ C {\bf 72}, 2267 (2012),
         [\href{http://xxx.lanl.gov/abs/1212.6017}
  {{\tt arXiv:1212.6017}}].
 
\bibitem{Bohmer:2011si} 
  C.~G.~Boehmer, T.~Harko and F.~S.~N.~Lobo,
  {\it{Wormhole geometries in modified teleparralel gravity and the energy conditions}},
  Phys.\ Rev.\ D {\bf 85}, 044033 (2012)
  [\href{http://xxx.lanl.gov/abs/1110.5756}
  {{\tt arXiv:1110.5756}}].
 
 
 
 
 
\bibitem{Ong:2013qja}
  Y.~C.~Ong, K.~Izumi, J.~M.~Nester and P.~Chen,
    {\it{Problems with Propagation and Time Evolution in f(T) Gravity}},
  Phys.\ Rev.\ D {\bf 88} (2013) 2,  024019,
     [\href{http://xxx.lanl.gov/abs/1303.0993}
 {{\tt arXiv:1303.0993}}].

\bibitem{Amoros:2013nxa}
  J.~Amoros, J.~de Haro and S.~D.~Odintsov,
    {\it{Bouncing Loop Quantum Cosmology from $F(T)$ gravity}},
     Phys.\ Rev.\ D {\bf 87}, 104037 (2013),
    [\href{http://xxx.lanl.gov/abs/1305.2344}
 {{\tt arXiv:1305.2344}}].

\bibitem{Nesseris:2013jea}
  S.~Nesseris, S.~Basilakos, E.~N.~Saridakis and L.~Perivolaropoulos,
  {\it{Viable f(T) models are practically indistinguishable from LCDM}},
  Phys.\ Rev.\ D {\bf 88}, 103010 (2013),
       [\href{http://xxx.lanl.gov/abs/1308.6142}
 {{\tt arXiv:1308.6142}}].

\bibitem{Bamba:2013ooa}
  K.~Bamba, S.~Capozziello, M.~De Laurentis, S.~'i.~Nojiri and D.~S\'{a}ez-G\'{o}mez,
     {\it{No further gravitational wave modes in $F(T)$ gravity}},
  Phys.\ Lett.\ B {\bf 727}, 194 (2013),
         [\href{http://xxx.lanl.gov/abs/1309.2698}
 {{\tt arXiv:1309.2698}}].

 
 
\bibitem{Basilakos:2013rua}
  S.~Basilakos, S.~Capozziello, M.~De Laurentis, A.~Paliathanasis and
M.~Tsamparlis,
     {\it{Noether symmetries and analytical solutions in f(T)-cosmology: A
 complete study}},
  Phys.\ Rev.\ D {\bf 88}, 103526 (2013),
           [\href{http://xxx.lanl.gov/abs/1311.2173}
 {{\tt arXiv:1311.2173}}].

 
 
\bibitem{Otalora:2014aoa} 
  G.~Otalora,
    {\it{A novel teleparallel dark energy model}},
            [\href{http://xxx.lanl.gov/abs/1402.2256 }
 {{\tt arXiv:1402.2256 }}].


 

  
\bibitem{Paliathanasis:2014iva}
  A.~Paliathanasis, S.~Basilakos, E.~N.~Saridakis, S.~Capozziello,
K.~Atazadeh, F.~Darabi and M.~Tsamparlis,
     {\it{New Schwarzschild-like solutions in f(T) gravity through Noether
 symmetries}},
   Phys.\ Rev.\ D {\bf 89}, 104042 (2014),
               [\href{http://xxx.lanl.gov/abs/1402.5935 }
 {{\tt arXiv:1402.5935 }}].


  
 \bibitem{Nashed:2014uta}
  G.~G.~L.~Nashed,
    {\it{$f(T)$ gravity theories and local Lorentz transformation}},
              [\href{http://xxx.lanl.gov/abs/1403.6937}
{{\tt arXiv:1403.6937}}].


 

\bibitem{Bengochea001}
    G.~R.~Bengochea,
   {\it{Observational information for f(T) theories and Dark Torsion}},
  Phys.\ Lett.\  {\bf B695}, 405 (2011),
[\href{http://xxx.lanl.gov/abs/1008.3188}
{{\tt arXiv:1008.3188}}].


\bibitem{Wang:2011xf}
    T.~Wang,
  {\it{Static Solutions with Spherical Symmetry in f(T) Theories}},
  Phys.\ Rev.\  {\bf D84}, 024042 (2011),
[\href{http://xxx.lanl.gov/abs/1102.4410}
{{\tt arXiv:1102.4410}}].

\bibitem{Miao003}
  R.~-X.~Miao, M.~Li and Y.~-G.~Miao,
  {\it{Violation of the first law of black hole thermodynamics in $f(T)$
 gravity}},
  JCAP {\bf 1111}, 033 (2011),
  [\href{http://xxx.lanl.gov/abs/1107.0515}
{{\tt arXiv:1107.0515}}].


\bibitem{Boehmer:2011gw}
  C.~G.~Boehmer, A.~Mussa and N.~Tamanini,
     {\it{Existence of relativistic stars in f(T) gravity}},
  Class.\ Quant.\ Grav.\  {\bf 28}, 245020 (2011),
[\href{http://xxx.lanl.gov/abs/1107.4455}
{{\tt arXiv:1107.4455}}].
  
\bibitem{Daouda001}
  M.~H. Daouda, M.~E.~Rodrigues and M.~J.~S.~Houndjo,
  {\it{New Static Solutions in f(T) Theory}},
  Eur.\ Phys.\ J.\ C {\bf 71}, 1817 (2011),
  [\href{http://xxx.lanl.gov/abs/1108.2920}
{{\tt arXiv:1108.2920}}].


 
  
\bibitem{Ferraro:2011ks}
  R.~Ferraro, F.~Fiorini,
 {\it {Spherically symmetric static spacetimes in vacuum f(T) gravity}},
  Phys.\ Rev.\ D {\bf 84}, 083518 (2011),
  [\href{http://xxx.lanl.gov/abs/1109.4209}
{{\tt arXiv:1109.4209}}].


\bibitem{Gonzalez:2011dr}
  P.~A.~Gonzalez, E.~N.~Saridakis and Y.~Vasquez,
      {\it{Circularly symmetric solutions in three-dimensional
Teleparallel, f(T) and
  Maxwell-f(T) gravity}},
[\href{http://xxx.lanl.gov/abs/1110.4024}
{{\tt arXiv:1110.4024}}].


\bibitem{Capozziello:2012zj}
  S.~Capozziello, P.~A.~Gonzalez, E.~N.~Saridakis and Y.~Vasquez,
    {\it{Exact charged black-hole solutions in D-dimensional f(T) gravity:
torsion vs curvature analysis}},
  JHEP {\bf 1302} (2013) 039,
    [\href{http://xxx.lanl.gov/abs/1210.1098}
{{\tt arXiv:1210.1098}}].



  
\bibitem{Atazadeh:2012am}
  K.~Atazadeh and M.~Mousavi,
 {\it{Vacuum spherically symmetric solutions in $f(T)$ gravity}},
  Eur.\ Phys.\ J.\ C {\bf 72}, 2272 (2012),
      [\href{http://xxx.lanl.gov/abs/1212.3764}
{{\tt arXiv:1212.3764}}].
 

  
\bibitem{Kofinas:2014owa}
  G.~Kofinas and E.~N.~Saridakis,
  {\it{Teleparallel equivalent of Gauss-Bonnet gravity and its
modifications}},
     [\href{http://xxx.lanl.gov/abs/1404.2249}
{{\tt arXiv:1404.2249}}].
 
  
\bibitem{Kofinas:2014aka} 
  G.~Kofinas, G.~Leon and E.~N.~Saridakis,
 {\it{Dynamical behavior in $f(T,T_G)$ cosmology}},
  Class.\ Quant.\ Grav.\  {\bf 31}, 175011 (2014),
       [\href{http://xxx.lanl.gov/abs/1404.7100}
{{\tt arXiv:1404.7100}}].

 

 

  \bibitem{WC1WC2} 
  Z.~Haghani, T.~Harko, H.~R.~Sepangi and S.~Shahidi,
 {\it{Weyl-Cartan-Weitzenboeck gravity as a generalization of teleparallel gravity}},
  JCAP {\bf 1210}, 061 (2012),
         [\href{http://xxx.lanl.gov/abs/1202.1879}
{{\tt arXiv:1202.1879}}].

 

   \bibitem{WC2}
  Z.~Haghani, T.~Harko, H.~R.~Sepangi and S.~Shahidi,
 {\it{Weyl-Cartan-Weitzenböck gravity through Lagrange multiplier}},
  Phys.\ Rev.\ D {\bf 88},   044024 (2013),
          [\href{http://xxx.lanl.gov/abs/1307.2229}
{{\tt arXiv:1307.2229}}].

 

\bibitem{Uzan:1999ch}
  J.~-P.~Uzan,
 {\it{Cosmological scaling solutions of nonminimally coupled scalar
 fields}},
  Phys.\ Rev.\ D {\bf 59}, 123510 (1999),
  [\href{http://xxx.lanl.gov/abs/gr-qc/9903004}
 {{\tt arXiv:gr-qc/9903004}}].


\bibitem{deRitis:1999zn}
  R.~de Ritis, A.~A.~Marino, C.~Rubano and P.~Scudellaro,
  {\it{Tracker fields from nonminimally coupled theory}},
  Phys.\ Rev.\ D {\bf 62}, 043506 (2000),
    [\href{http://xxx.lanl.gov/abs/hep-th/9907198}
 {{\tt arXiv:hep-th/9907198}}].


\bibitem{Bertolami:1999dp}
  O.~Bertolami and P.~J.~Martins,
   {\it{Nonminimal coupling and quintessence}},
  Phys.\ Rev.\ D {\bf 61}, 064007 (2000),
      [\href{http://xxx.lanl.gov/abs/gr-qc/9910056}
 {{\tt arXiv:gr-qc/9910056}}].


\bibitem{Faraoni:2000wk}
  V.~Faraoni,
   {\it{Inflation and quintessence with nonminimal coupling}},
  Phys.\ Rev.\ D {\bf 62}, 023504 (2000),
  [\href{http://xxx.lanl.gov/abs/gr-qc/0002091}
 {{\tt arXiv:gr-qc/0002091}}].

 




\bibitem{Amendola:1993uh}
  L.~Amendola,
  {\it{Cosmology with nonminimal derivative couplings}},
  Phys.\ Lett.\  B {\bf 301}, 175 (1993),
       [\href{http://xxx.lanl.gov/abs/gr-qc/9302010}
 {{\tt arXiv:gr-qc/9302010}}].


\bibitem{Capozziello:1999xt}
  S.~Capozziello, G.~Lambiase and H.~J.~Schmidt,
   {\it{Nonminimal derivative couplings and inflation in generalized theories
 of gravity}},
  Annalen Phys.\  {\bf 9}, 39 (2000),
  [\href{http://xxx.lanl.gov/abs/gr-qc/9906051}
 {{\tt arXiv:gr-qc/9906051}}].


\bibitem{Daniel:2007kk}
  S.~F.~Daniel and R.~R.~Caldwell,
   {\it{Consequences of a cosmic scalar with kinetic coupling to curvature}},
  Class.\ Quant.\ Grav.\  {\bf 24}, 5573 (2007),
    [\href{http://xxx.lanl.gov/abs/0709.0009}
 {{\tt arXiv:0709.0009}}].


\bibitem{Saridakis:2010mf}
  E.~N.~Saridakis and S.~V.~Sushkov,
   {\it{Quintessence and phantom cosmology with non-minimal derivative
  coupling}},
  Phys.\ Rev.\ D {\bf 81}, 083510 (2010),
     [\href{http://xxx.lanl.gov/abs/1002.3478}
 {{\tt arXiv:1002.3478}}].


\bibitem{Sadjadi:2010bz}
  H.~M.~Sadjadi,
  {\it{Super-acceleration in non-minimal derivative coupling model}},
  Phys.\ Rev.\ D {\bf 83}, 107301 (2011),
  [\href{http://xxx.lanl.gov/abs/1012.5719}
 {{\tt arXiv:1012.5719}}].

  
  


\bibitem{ArmendarizPicon:2000ah}
  C.~Armendariz-Picon, V.~F.~Mukhanov and P.~J.~Steinhardt,
  {\it{Essentials of k essence}},
  Phys.\ Rev.\ D {\bf 63}, 103510 (2001),
    [\href{http://xxx.lanl.gov/abs/astro-ph/0006373}
 {{\tt arXiv:astro-ph/0006373}}].


 



 \bibitem{Horndeski}
 G.~W.~Horndeski,
 {\it{Second-order scalar-tensor field equations in a four-dimensional
 space}},
Int.\ J.\ Theor.\ Phys.\ 10,
363-384 (1974).




\bibitem{DeFelice:2010nf}
  A.~De Felice and S.~Tsujikawa,
   {\it{Generalized Galileon cosmology}},
  Phys.\ Rev.\ D {\bf 84}, 124029 (2011),
 [\href{http://xxx.lanl.gov/abs/1008.4236}
 {{\tt arXiv:1008.4236}}].

\bibitem{Deffayet:2011gz}
  C.~Deffayet, X.~Gao, D.~A.~Steer and G.~Zahariade,
 {\it{From k-essence to generalised Galileons}},
  Phys.\ Rev.\ D {\bf 84}, 064039 (2011),
 [\href{http://xxx.lanl.gov/abs/1103.3260}
 {{\tt arXiv:1103.3260}}].


\bibitem{DeFelice:2011bh}
  A.~De Felice and S.~Tsujikawa,
   {\it{Conditions for the cosmological viability of the most general
 scalar-tensor theories and their applications to extended Galileon dark
 energy models}},
  JCAP {\bf 1202}, 007 (2012),
 [\href{http://xxx.lanl.gov/abs/1110.3878}
 {{\tt arXiv:1110.3878}}].


 



\bibitem{Bertolami:2007gv} 
  O.~Bertolami, C.~G.~Boehmer, T.~Harko and F.~S.~N.~Lobo,
   {\it{Extra force in f(R) modified theories of gravity}},
  Phys.\ Rev.\ D {\bf 75}, 104016 (2007),
   [\href{http://xxx.lanl.gov/abs/0704.1733}
 {{\tt arXiv:0704.1733}}].

 
 
 

    
\bibitem{Bertolami:2008zh} 
  O.~Bertolami, J.~Paramos, T.~Harko and F.~S.~N.~Lobo,
    {\it{Non-minimal curvature-matter couplings in modified gravity}},
       [\href{http://xxx.lanl.gov/abs/0811.2876}
 {{\tt arXiv:0811.2876}}].

 
 
 
    
\bibitem{Bertolami:2008ab} 
  O.~Bertolami, F.~S.~N.~Lobo and J.~Paramos,
    {\it{Non-minimum coupling of perfect fluids to curvature}},
  Phys.\ Rev.\ D {\bf 78}, 064036 (2008),
         [\href{http://xxx.lanl.gov/abs/0806.4434}
 {{\tt arXiv:0806.4434}}].

 
 
 
  
\bibitem{Bertolami:2009ic} 
  O.~Bertolami and J.~Paramos,
    {\it{Mimicking dark matter through a non-minimal gravitational coupling with matter}},
  JCAP {\bf 1003}, 009 (2010),
           [\href{http://xxx.lanl.gov/abs/0906.4757}
 {{\tt arXiv:0906.4757}}].

  
  
  

\bibitem{Harko:2008qz}
  T.~Harko,
   {\it{Modified gravity with arbitrary coupling between matter and geometry}},
  Phys.\ Lett.\ B {\bf 669}, 376 (2008),
             [\href{http://xxx.lanl.gov/abs/0810.0742}
 {{\tt arXiv:0810.0742}}].

  
  
 

  
  
\bibitem{Harko:2010mv}
  T.~Harko and F.~S.~N.~Lobo,
  {\it{f(R,$L_{m}$) gravity}},
  Eur.\ Phys.\ J.\ C {\bf 70}, 373 (2010),
               [\href{http://xxx.lanl.gov/abs/1008.4193}
 {{\tt arXiv:1008.4193}}].

  
  
  
 
  
  
\bibitem{Harko:2012hm}
  T.~Harko, F.~S.~N.~Lobo and O.~Minazzoli,
   {\it{Extended $f(R,L_m)$ gravity with generalized scalar field and kinetic
 term dependences}},
  Phys.\ Rev.\ D {\bf 87},   047501 (2013),
                [\href{http://xxx.lanl.gov/abs/1210.4218}
 {{\tt arXiv:1210.4218}}].

  
 
  
\bibitem{Wang:2012rw}
  J.~Wang and K.~Liao,
   {\it{Energy conditions in f(R, L(m)) gravity}},
  Class.\ Quant.\ Grav.\  {\bf 29}, 215016 (2012),
                  [\href{http://xxx.lanl.gov/abs/1212.4656}
 {{\tt arXiv:1212.4656}}].

\bibitem{Harko:2014gwa} 
  T.~Harko and F.~S.~N.~Lobo,
  {\it{Generalized curvature-matter couplings in modified gravity}},
  Galaxies 2 (2014) 3, 410-465
    [\href{http://xxx.lanl.gov/abs/1407.2013}
  {{\tt arXiv:1407.2013}}].
  
 

\bibitem{Harko:2011kv}
   T.~Harko, F.~S.~N.~Lobo, S.~Nojiri and S.~D.~Odintsov,
   {\it{$f(R,T)$ gravity}},
  Phys.\ Rev.\ D {\bf 84}, 024020 (2011),
                    [\href{http://xxx.lanl.gov/abs/1104.2669}
 {{\tt arXiv:1104.2669}}].


 
 
  
   

\bibitem{Momeni:2011am}
  M.~Jamil, D.~Momeni, M.~Raza and R.~Myrzakulov,
   {\it{Reconstruction of some cosmological models in f(R,T) gravity}},
  Eur.\ Phys.\ J.\ C {\bf 72}, 1999 (2012),
                      [\href{http://xxx.lanl.gov/abs/1107.5807}
 {{\tt arXiv:1107.5807}}].

 
 
  
  
\bibitem{Sharif:2012zzd}
  M.~Sharif and M.~Zubair,
    {\it{Thermodynamics in f(R,T) Theory of Gravity}},
  JCAP {\bf 1203}, 028 (2012),
                        [\href{http://xxx.lanl.gov/abs/1204.0848}
 {{\tt arXiv:1204.0848}}].

 
 
  
  
\bibitem{Alvarenga:2013syu}
  F.~G.~Alvarenga, A.~de la Cruz-Dombriz, M.~J.~S.~Houndjo, M.~E.~Rodrigues
and D.~Sáez-Gómez,
     {\it{Dynamics of scalar perturbations in f(R,T) gravity}},
  Phys.\ Rev.\ D {\bf 87}, 103526 (2013),
     [\href{http://xxx.lanl.gov/abs/1302.1866}
 {{\tt arXiv:1302.1866}}].

 
 
  
  
\bibitem{Shabani:2013djy}
  H.~Shabani and M.~Farhoudi,
    {\it{f(R,T) Cosmological Models in Phase Space}},
  Phys.\ Rev.\ D {\bf 88}, 044048 (2013),
         [\href{http://xxx.lanl.gov/abs/1306.3164}
 {{\tt arXiv:1306.3164}}].

 
 
  

\bibitem{fRT} 
  Z.~Haghani, T.~Harko, F.~S.~N.~Lobo, H.~R.~Sepangi and S.~Shahidi,
 {\it{Further matters in space-time geometry: f(R,T,$R_{\mu\nu}T^{\mu\nu}$) gravity}},
  Phys.\ Rev.\ D {\bf 88}, no. 4, 044023 (2013),
           [\href{http://xxx.lanl.gov/abs/1304.5957}
 {{\tt arXiv:1304.5957}}].

 
 

\bibitem{Odintsov:2013iba} 
  S.~D.~Odintsov and D.~Sáez-Gómez,
 {\it{$f(R, T, R_{\mu\nu} T^{\mu\nu})$ gravity phenomenology and $\Lambda$CDM universe}},
  Phys.\ Lett.\ B {\bf 725}, 437 (2013),
             [\href{http://xxx.lanl.gov/abs/1304.5411}
 {{\tt arXiv:1304.5411}}].
 
  
 
\bibitem{Geng:2011aj}
  C.~-Q.~Geng, C.~-C.~Lee, E.~N.~Saridakis and Y.~-P.~Wu,
   {\it{'Teleparallel' Dark Energy}},
  Phys.\ Lett.\ B {\bf 704}, 384 (2011),
  [\href{http://xxx.lanl.gov/abs/1109.1092}
{{\tt arXiv:1109.1092}}].

\bibitem{Wei:2011yr}
    H.~Wei,
   {\it{Dynamics of Teleparallel Dark Energy}},
  Phys.\ Lett.\ B {\bf 712}, 430 (2012);
    [\href{http://xxx.lanl.gov/abs/1109.6107}
{{\tt arXiv:1109.6107}}].

  
\bibitem{Geng:2011ka}
  C.~-Q.~Geng, C.~-C.~Lee, E.~N.~Saridakis,
 {\it{Observational Constraints on Teleparallel Dark Energy}},
  JCAP {\bf 1201}, 002 (2012),
 [\href{http://xxx.lanl.gov/abs/1110.0913}
 {{\tt arXiv:1110.0913}}].


 \bibitem{Xu:2012jf}
   C.~Xu, E.~N.~Saridakis and G.~Leon,
  {\it{Phase-Space analysis of Teleparallel Dark Energy}},
  JCAP {\bf 1207}, 005 (2012),
     [\href{http://xxx.lanl.gov/abs/1202.3781}
 {{\tt arXiv:1202.3781}}].

  
 \bibitem{Otalora:2013dsa}
  G.~Otalora,
    {\it{Cosmological dynamics of tachyonic teleparallel dark energy}},
  Phys.\ Rev.\ D {\bf 88}, 063505 (2013),
          [\href{http://xxx.lanl.gov/abs/1305.5896}
{{\tt arXiv:1305.5896}}].


 \bibitem{Geng:2013uga}
  C.~-Q.~Geng, J.~-A.~Gu and C.~-C.~Lee,
   {\it{Singularity Problem in Teleparallel Dark Energy Models}},
   Phys.\ Rev.\ D {\bf 88}, 024030 (2013),
        [\href{http://xxx.lanl.gov/abs/1306.0333}
{{\tt arXiv:1306.0333}}].



\bibitem{Otalora:2013tba} 
  G.~Otalora,
   {\it{Scaling attractors in interacting teleparallel dark energy}},
  JCAP {\bf 1307}, 044 (2013),
          [\href{http://xxx.lanl.gov/abs/1305.0474}
{{\tt arXiv:1305.0474}}].


 



\bibitem{Sadjadi:2013nb} 
  H.~M.~Sadjadi,
    {\it{Notes on teleparallel cosmology with nonminimally coupled scalar field}},
  Phys.\ Rev.\ D {\bf 87}, 064028 (2013),
            [\href{http://xxx.lanl.gov/abs/1302.1180}
{{\tt arXiv:1302.1180}}].

 
  
\bibitem{Kucukakca:2013mya} 
  Y.~Kucukakca,
   {\it{Scalar tensor teleparallel dark gravity via Noether symmetry}},
  Eur.\ Phys.\ J.\ C {\bf 73}, 2327 (2013),
              [\href{http://xxx.lanl.gov/abs/1404.7315}
{{\tt arXiv:1404.7315}}].

 
    
\bibitem{Harko:2014sja} 
  T.~Harko, F.~S.~N.~Lobo, G.~Otalora and E.~N.~Saridakis,
   {\it{Nonminimal torsion-matter coupling extension of f(T) gravity}},
  Phys.\ Rev.\ D {\bf 89}, 124036 (2014),
      [\href{http://xxx.lanl.gov/abs/1404.6212}
{{\tt arXiv:1404.6212}}].
 


\bibitem{Kiani:2013pba}
  F.~Kiani and K.~Nozari,
    {\it{Energy conditions in $F(T,\Theta)$ gravity and compatibility with a stable de Sitter
  solution}},
  Phys.\ Lett.\ B {\bf 728}, 554 (2014),
        [\href{http://xxx.lanl.gov/abs/1309.1948}
{{\tt arXiv:1309.1948}}].
 
 
 

\bibitem{Weitzenb23}
  Weitzenb\"{o}ck R.,
  \emph{Invarianten Theorie},
  Nordhoff, Groningen (1923).


\bibitem{Mukhanov:1990me}
  V.~F.~Mukhanov, H.~A.~Feldman and R.~H.~Brandenberger,
     {\it{Theory of cosmological perturbations. Part 1. Classical perturbations.
 Part 2. Quantum theory of perturbations. Part 3. Extensions}},
  Phys.\ Rept.\  {\bf 215}, 203 (1992).


\bibitem{DeFelice:2010pv}
  A.~De Felice and S.~Tsujikawa,
     {\it{Cosmology of a covariant Galileon field}},
  Phys.\ Rev.\ Lett.\  {\bf 105}, 111301 (2010),
          [\href{http://xxx.lanl.gov/abs/1007.2700}
{{\tt arXiv:1007.2700}}].
 
  
 
\bibitem{Bogdanos:2009uj}
  C.~Bogdanos and E.~N.~Saridakis,
  {\it{Perturbative instabilities in Horava gravity}},
  Class.\ Quant.\ Grav.\  {\bf 27}, 075005 (2010),
            [\href{http://xxx.lanl.gov/abs/0907.1636}
{{\tt arXiv:0907.1636}}].
 
  

\bibitem{Wang:2009yz} 
  A.~Wang and R.~Maartens,
   {\it{Linear perturbations of cosmological models in the Horava-Lifshitz theory of gravity without detailed balance}},
  Phys.\ Rev.\ D {\bf 81}, 024009 (2010),
              [\href{http://xxx.lanl.gov/abs/0907.1748}
{{\tt arXiv:0907.1748}}].
 
 
 

\bibitem{Dent:2013awa}
  J.~B.~Dent, S.~Dutta, E.~N.~Saridakis and J.~-Q.~Xia,
   {\it{Cosmology with non-minimal derivative couplings:perturbation analysis
 and observational constraints}},
  JCAP {\bf 1311}, 058 (2013),
         [\href{http://xxx.lanl.gov/abs/1309.4746}
{{\tt arXiv:1309.4746}}].
 
 
  


  \bibitem{Planck} 
  P.~A.~R.~Ade {\it et al.}  [Planck Collaboration],
    {\it{Planck 2013 results. XXII. Constraints on inflation}},
             [\href{http://xxx.lanl.gov/abs/1303.5082}
{{\tt arXiv:1303.5082}}].
 
 
 
\bibitem{BICEP2}
  P.~A.~R.~Ade {\it et al.}  [BICEP2 Collaboration],
   {\it{Detection of B-Mode Polarization at Degree Angular Scales by BICEP2}},
  Phys.\ Rev.\ Lett.\  {\bf 112}, 241101 (2014),
               [\href{http://xxx.lanl.gov/abs/1403.3985}
{{\tt arXiv:1403.3985}}].
 
  
\end{thebibliography}
\end{document}